\documentclass[
 reprint,
frontmatterverbose, 
nofootinbib,
 amsmath,amssymb,
 aps,
prd,
floatfix,
]{revtex4-2}

\usepackage{graphicx}

\usepackage{dcolumn}
\usepackage{bm}
\usepackage{pifont}
\newcommand{\cmark}{\ding{51}}%
\newcommand{\xmark}{\ding{55}}%

\usepackage{graphics}

\usepackage{hyperref}
\usepackage{xcolor,colortbl}
\definecolor{Gray}{gray}{0.75}

\begin{document}

\preprint{}

\title{\texorpdfstring{Reflecting on Chirality: CP-violating extensions of the single scalar-leptoquark solutions for the $(g-2)_{e,\mu}$ puzzles and their implications for lepton EDMs}{Combining scalar leptoquark solutions to the (g-2) e and mu puzzle and their implications for d e or mu}}

\author{Innes Bigaran}
\email{innes.bigaran@unimelb.edu.au}
 \affiliation{School of Physics, The University of Melbourne, Victoria 3010, Australia}
\author{Raymond R. Volkas}
 \email{raymondv@unimelb.edu.au}
 \affiliation{School of Physics, The University of Melbourne, Victoria 3010, Australia}

\date{\today}
\begin{abstract}
We study the two scalar leptoquarks capable of generating chirally-enhanced, sign-dependent contributions to lepton magnetic dipole moments~(MDMs) and electric dipole moments~(EDMs), $R_2\sim (\mathbf{3}, \mathbf{2}, 7/6)$ and $S_1\sim (\mathbf{3}, \mathbf{1}, -1/3)$. We consider the case in which the electron and muon sectors are decoupled, and leptoquark couplings are assigned complex values. Adopting the coupling ansatz that the electron dipole operator is generated by charm-containing loops, and muon dipole operator by top-containing loops, we find that both minimal leptoquark models remain viable solutions for reconciling anomalies in the muon and electron MDMs, accounting for either of the two current (disparate) electron MDM results from Cs and Rb interferometry experiments. We also examine the correlated corrections to the muon and electron masses generated by these models, and argue that to minimise fine-tuning this introduces an upper bound on viable leptoquark ($\phi$) masses, $m_{\phi}<\mathcal{O}(4)$ TeV. Similar arguments allow us to make a prediction for the upper bound of the muon EDM generated by these models, $|d_\mu|< \mathcal{O}(10^{-22})\; e$ cm, which could be within reach of upcoming experimental programs, including Muon g$-$2 at Fermilab~(FNAL), and muEDM at Paul Scherrer Institut~(PSI).
\end{abstract}

\maketitle

\section{Introduction}
\label{sec:Intro}
Lepton magnetic (MDM) and electric (EDM) dipole moments are sensitive precision probes for physics beyond the Standard Model~(SM). In terms of the effective interactions of SM fields, the EDM~ ($d_\ell$) and MDM~($a_\ell$) enter via the following
\begin{equation}
\begin{aligned}
    \mathcal{L}_{\ell \gamma}&=  \overline{\ell} \left( a_\ell\frac{e}{4m_\ell} \sigma_{\mu\nu} -d_\ell\frac{i}{2} \sigma_{\mu\nu}\gamma_5\right)\ell F^{\mu\nu}, \label{agamma}
\end{aligned}
\end{equation}
where $F^{\mu\nu}= \partial^\mu A^\nu - \partial^\nu A^\mu$, and 
\begin{align}
    a_\ell \equiv \frac{1}{2} (g-2)_\ell,
\end{align}
where this represents the deviation of this observable from the tree-level value, $g= 2$.

An accurate determination of the MDMs of charged-leptons has long been an important test of the Standard Model~(SM). Given the precision of the SM predictions, any deviation from these could provide a `smoking gun' for BSM physics -- particularly where these models could generate loop-level, flavour-violating interactions.

A persistent deviation between experiment and prediction for the muon MDM has garnered significant attention in the literature. The most recent measurement of $a_\mu^\text{exp}$ from the Fermilab~(FNAL) Muon g$-$2 collaboration~\cite{Abi:2021gix} yields the following global average
\begin{equation}
\Delta a_\mu= a_\mu^{\text{exp}} - a_\mu^{\text{SM}}=(2.51\pm0.59) \times 10^{-9}, \label{g-2mu}
\end{equation}
when combined with the Brookhaven Muon g$-$2 results~\cite{Muong-2:2004fok}. The SM prediction is taken from the theory white paper~\cite{Aoyama:2020ynm}, the result of a combination of work in references~\cite{aoyama:2012wk, Aoyama:2019ryr,czarnecki:2002nt,gnendiger:2013pva, davier:2017zfy,keshavarzi:2018mgv,colangelo:2018mtw, hoferichter:2019gzf,davier:2019can,keshavarzi:2019abf,kurz:2014wya,melnikov:2003xd,masjuan:2017tvw,Colangelo:2017fiz,hoferichter:2018kwz,gerardin:2019vio,bijnens:2019ghy,colangelo:2019uex,Blum:2019ugy,colangelo:2014qya}. This represents $4.2\,\sigma$ evidence for the existence of BSM physics.

In contrast to this measurement of the muon, the MDM of the electron can be indirectly probed using spectroscopic measurements of the fine structure constant, $\alpha_{em}$, as input for the SM theory calculation. The result from the most recent measurement in caesium~(Cs) suggests an anomaly in the electron MDM
\begin{equation}
\Delta a_e^\text{Cs}=a_e^{\text{exp}} - a_e^{\text{SM, Cs}}= -(8.8\pm3.6) \times 10^{-13},\label{g-2eCs}
\end{equation}
where this corresponds to a $2.5\,\sigma$ deviation from the SM~\cite{articleParker}.
In contrast to this, the result from the most recent measurement in rubidium~(Rb) yields
\begin{equation}
\Delta a^\text{Rb}_e=a_e^{\text{exp}} - a_e^{\text{SM, Rb}}= (4.8\pm3.0) \times 10^{-13}.\label{g-2eRb}
\end{equation}
which suggests a $1.6\,\sigma$ deviation~\cite{Morel:2020dww}. These determinations are not consistent with each other within the stated errors, so one must conclude that the electron MDM situation requires clarification. For the moment, we thus treat these two results as independent cases and remain agnostic about where the resolution will lie.

It is important to note that were the result of equation~\eqref{g-2eCs} to be confirmed, then $\Delta a_\mu$ and $\Delta a_e$ would be of \emph{opposite sign}, which would be an interesting difficulty to be overcome when searching for a common explanation. Such considerations motivate a search for flavour-violating BSM effects to achieve a common solution for these two (putative) anomalies.

Leptonic EDMs add to the MDM information by providing a probe of CP-violating effects beyond the SM. Non-zero permanent EDMs of leptons would break parity (P) and time-reversal (T) symmetry and, by the CPT theorem, also CP symmetry. The only verified source of CP violation in the SM is the non-zero phase of the CKM matrix, which predicts EDMs that are orders of magnitude below current experimental bounds. Therefore, a measurement of a relatively sizeable leptonic EDMs at future experiments would necessitate an authentic BSM source of CP violation.

Given the anomaly or anomalies in the MDMs, we may speculate as to whether this tension with the SM could also manifest in the EDMs $d_e$ and $d_\mu$ in anticipation of potential future measurements. The current bounds for the muon~\cite{Bennett:2008dy} and the electron~\cite{Andreev:2018ayy}are given by
\begin{equation}
\begin{aligned}
       |d_\mu|&< 1.5 \times 10^{-19} \;e\;\text{cm}\;\;\;\; 90\% \; \text{CL}\;,\\
       |d_e|&< 1.1 \times 10^{-29} \;e\;\text{cm}\;\;\;\; 90\% \; \text{CL}\;.
\end{aligned}
\end{equation}

In reference~\cite{Bigaran:2020jil}, we discussed the viability of models of single-scalar leptoquarks~(LQs) to simultaneously ameliorate the anomaly in $\Delta a_\mu$, as it stood prior to the FNAL result quoted in equation~\eqref{g-2mu},  and the potential anomaly $\Delta a_e^{\rm Cs}$ of equation~\eqref{g-2eCs}. Contributions to $d_e$ and $d_\mu$ were rendered negligible in the parameter space explored, by setting the LQ couplings to solely real values. In the present work, we explore the implications of these established models in an extended parameter study where contributions to $d_e$ and $d_\mu$ exist. Lepton EDMs have been studied in the context of scalar LQ models, for example, in references~\cite{Crivellin:2018qmi,Dekens:2018bci,Fuyuto:2018scm, Bernreuther:2021elu, Lee:2021jdr}. 

The outline of this paper is as follows; in Section~\ref{sec:SLQ} we introduce scalar leptoquark models, comment on their manifestation in leptonic MDMs and EDMs, and outline our LQ coupling flavour ansatz. In Section~\ref{sec:LeptonMasses}, we comment on the constraints on these models from charged-lepton mass corrections in the context of fine-tuning, and present the implications of these constraints on LQ masses and predictions for the generated EDMs. In Section~\ref{sec:Additional Constraints}, we present an overview of important additional constraints, and illustrate the viable parameter space for both LQ models. Finally, in Section~\ref{sec:ThirdFamily} we comment briefly on the implications of these models on the tau lepton sector, before concluding in Section~\ref{sec:conclusion}.

\section{Scalar Leptoquarks and \emph{Getting Chirality Right}}
\label{sec:SLQ}
 
 Letoquarks are hypothetical exotic particles that directly Yukawa couple to SM leptons and quarks. Motivated by the introduction of direct lepton-quark couplings, rather than separately considering lepton~(L) or baryon~(B) number conservation, these are absorbed into the definition of a new conserved quantity~\cite{articlePSFN} \emph{fermion} number: $F=3B+L.$ Even though vector LQs have also been considered in the context of magnetic moment anomalies, we would expect vector LQ to be associated with gauge-symmetry extensions, complicating the construction of a full UV-complete model. Therefore, in the interest of simplicity, we focus only on scalar LQs in this work. Table~\ref{table:LQtransform} gives an overview of the finite set of scalar LQs, including their gauge-group transformations and fermion numbers, adopting the naming convention of reference~\cite{Dorsner:2016wpm}.
 
 The fermion number characterises the type of lepton-quark interactions; for generic charged-leptons~($\ell$) and quarks~($q$), $|F|=2$ LQs couple to fermion bilinears of the form $\ell q$, and $|F|=0$ couple to $\overline{\ell} q$. The Yukawa couplings can be expressed as
\begin{equation}
    \mathcal{L}_\ell = \overline{\ell^{(c)}}\; \left[y^{R}P_R + y^{L}P_L\right]\;q\;\phi^\dagger+h.c. \label{genericLQ}
\end{equation}
for general scalar LQ models. The coupling constants $y^{L}$ and $y^{R}$ are labelled by the chirality of the quark in the interaction. Leptoquarks which have both left-handed~(LH) and right-handed~(RH) couplings, namely $S_1$ and $R_2$,  are referred to as \emph{mixed-chiral}. We will see that the chirality of the LQ couplings is paramount for generating sign-dependence in contributions to $\Delta a_\ell$, relevant if equation~\eqref{g-2eCs} is correct. Additionally, a mixed-chiral LQ can yield enhanced contributions to the effective lepton dipole vertex, via a mass-insertion on an internal fermion leg.

  \begin{table}[t!]
\centering
 \renewcommand*{\arraystretch}{1.3}
\begin{tabular}{ |c|c|c|c| }
 \hline
 Symbol & $SU(3)_C \otimes SU(2)_L \otimes U(1)_Y$ &  Mixed chiral & $|F|$\\
 \hline
$\tilde{S}_1$   & $ (\mathbf{3}, \mathbf{1}, -4/3)$  & \xmark & 2\\
 \rowcolor{Gray}
$S_1$   & $ (\mathbf{3}, \mathbf{1}, -1/3)$  & \cmark  & 2\\
$S_3$&   $ (\mathbf{3}, \mathbf{3}, -1/3)$ & \xmark & 2\\
$\overline{S}_1$   & $ (\mathbf{3}, \mathbf{1}, 2/3)$ & \xmark & 2\\
\rowcolor{Gray}
${R}_2$&   $ (\mathbf{3}, \mathbf{2}, 7/6)$ & \cmark & 0\\
$\tilde{R}_2$&   $ (\mathbf{3}, \mathbf{2}, 1/6)$ & \xmark & 0\\
 \hline
\end{tabular}
\caption{Scalar LQs and their transformation properties, under the hypercharge convention $Q=I_3+Y$. The second-last column indicates whether the LQ has mixed-chiral couplings.}
\label{table:LQtransform}
 \end{table}
 
 Note that the $|F|=2$ scalar leptoquarks can also have gauge-invariant di-quark couplings, leading to a violation of baryon number conservation and hence strong constraints from proton decay. We thus impose global conservation of baryon number throughout to forbid diquark couplings of $S_1$.

 \subsection{Scalar LQ contributions to MDMs}
 \label{sec:SLQ_sub:MDM}
 We comment briefly on this calculation here, but refer the reader to reference~\cite{Bigaran:2020jil} for a more extended discussion.
 
Following from equation~\eqref{agamma}, the following Lagrangian parameterises the contributions to $a_\ell$
\begin{equation}
\begin{aligned}
    \mathcal{L}_{\ell \gamma}&\supset  e\overline{\ell}\frac{a_\ell}{4m_\ell} \sigma_{\mu\nu}F^{\mu\nu}\ell,\\
    &\supset \frac{1}{2}  e\overline{\ell}  \sigma_{\mu\nu}F^{\mu\nu}\left( \sigma_L^\ell P_L+ \sigma_R^\ell P_R\right)\ell. \label{generical}
  \end{aligned}
\end{equation}
where $\sigma_{L/R}^\ell$ parameterise the effective left- and right-chiral interactions, as $P_{L(R)}= \frac{1}{2}(1\mp \gamma_5)$. We also note that $\sigma_{L/R}^\ell$ are not independent: in fact, $\sigma_{L}^\ell=[\sigma_{R}^\ell]^*$. Equation~\eqref{generical} reveals, via coefficient matching\footnote{Here have defined $\sigma_{L/R}^\ell= i[\sigma_{L/R}^\ell]_{2002.12544}$, \emph{i.e.} in contrast with the definitions used in reference~\cite{Bigaran:2020jil}. This is to make the discussion of EDMs clearer, in that imaginary-valued LQ couplings correspond directly to imaginary-valued effective interactions. }, that
\begin{equation}
\Delta a_\ell=  m_\ell (\sigma_L^\ell+\sigma_R^\ell)= 2 m_\ell \text{Re}(\sigma_L^\ell) ,\label{core1}
\end{equation}

\begin{figure}[t!]
\centering
\includegraphics[scale=1.2]{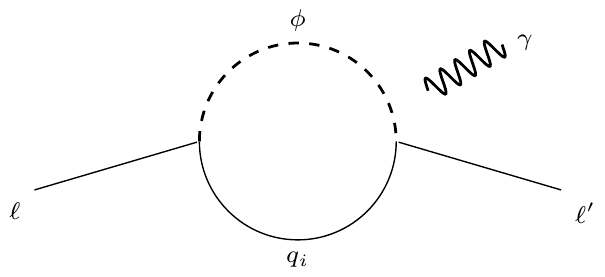}
     \caption{ Dominant contributions to the $\ell \to \ell' \gamma$ processes from scalar LQs -- including $d_\ell$ and $a_\ell$ corrections, where $\ell=\ell'$. The photon line could be connected to either of the internal legs.}  \label{feyndi}
    \end{figure}

The leading-order topologies for these corrections are illustrated in Figure~\ref{feyndi}. Their contributions to $\sigma_{L/R}^\ell$ are well-established in the literature~\cite{Dorsner:2016wpm}. For $F=0$ scalar LQs~(\emph{e.g.} $R_2$), via the Lagrangian in equation~\eqref{genericLQ}, we have that
\begin{equation}
\begin{aligned}
    \sigma^{\ell}_L 
    = -\frac{N_c}{16\pi^2 m_{\phi}^2}
    \sum_{q} \large[m_\ell(|y^R_{\ell q}|^2&+|y^L_{\ell q}|^2)\kappa\\
    &+y^{R}_{\ell q} y^{L*}_{\ell q} m_q \kappa'\large],\label{sigmaL}
\end{aligned}
\end{equation}where
\begin{equation}
\begin{aligned}
&\kappa(x_q)= Q_{\phi}f_S(x_q)-f_F(x_q),\;\\  &\kappa'(x_q)= Q_{\phi}g_S(x_q)-g_F(x_q),\;\label{kappas}
\end{aligned}
\end{equation}
with $x_q=m_q^2/m_\phi^2$.  These contributions are proportional to the number of colours, $N_c=3$, and are summed over quark flavours $q$ running in the loop. The electric charge of the field $\phi$ is given by $Q_{\phi}$, and the loop functions in equation~\eqref{kappas} are given by~\cite{Dorsner:2016wpm}
{\small
\begin{equation}
\begin{aligned}
f_S(x)&= \frac{x+1}{4(x-1)^2}- \frac{x\log(x)}{2(x-1)^3},\\
f_F(x)&= \frac{x^2-5x-2}{12(x-1)^3}+ \frac{x\log(x)}{2(x-1)^4},\\
g_S(x)&=\frac{1}{x-1} -\frac{\log(x)}{(x-1)^2},\\
g_F(x)&= \frac{x-3}{2(x-1)^2}+\frac{\log(x)}{(x-1)^3}.
\end{aligned}\label{loopfun}
\end{equation}
}

\noindent For $|F|=2$ LQs~(\emph{e.g.} $S_1$), the above effective contributions to the MDM or EDM are calculated similarly, but with $Q_\phi \mapsto -Q_\phi$. For a scalar LQ without mixed-chiral Yukawa couplings, the $y^R\, y^{L*}$ term in equation~\eqref{sigmaL} is not present and contributions from each propagator are of definite relative sign. However, for mixed-chiral scalar LQs, there are terms proportional to $\kappa'$, allowing us to vary the sign of the BSM contribution.

As summarised in Table~\ref{table:LQtransform}, by virtue of their chiral nature the $S_1$ and $R_2$ leptoquarks are able to induce sign-dependent $\Delta a_{e,\mu}$ contributions enhanced by heavy internal quark masses. These two LQs have also garnered recent attention in other flavour anomaly studies (see, for example, those of references~\cite{Becirevic:2016oho,Popov:2016fzr,ColuccioLeskow:2016dox,Buttazzo:2017ixm,Angelescu:2018tyl,Dorsner:2019itg, Bigaran:2019bqv,Crivellin:2019qnh,Bordone:2020lnb,Crivellin:2020mjs,Gherardi:2020qhc,Dash:2021suj,Greljo:2021xmg,Marzocca:2021miv, Marzocca:2021azj, Mohapatra:2021ynn,Murgui:2021bdy,Perez:2021ddi,Singirala:2021gok,Zhang:2021dgl,Angelescu:2021lln}). The relevant LQ couplings for each extension, represented here as $3\times 3$ Yukawa coupling matrices, are given by\footnote{We implement the interaction basis with neutrinos in their flavour eigenstates (\emph{i.e.} PMNS rotation matrix is set to the identity throughout) and consider only terms in which these fields couple as leptoquarks.}
\begin{equation}
 \begin{aligned}
 \mathcal{L}_{\text{int}}^{S_1} = & \left( \overline{L_L^c} \lambda_{LQ} Q_L +  \overline{e_R^c}\lambda_{eu} u_R\right) S_1^\dagger+ h.c.,
 \end{aligned}
 \end{equation}
  \begin{equation}
 \begin{aligned}
 \mathcal{L}_{\text{int}}^{R_2} &= \left( \overline{L_L}\lambda_{Lu} u_R +  \overline{e_R} \lambda_{eQ} Q_L \right)R_2^\dagger + h.c.
 \label{lagrange2}
 \end{aligned}
 \end{equation}
The doublet, $R_2$, can be expressed in terms of its electric charge-definite components $R_2 \sim (R_2^{5/3}, R_2^{2/3})$,
with charges as indicated by the superscripts. We assume negligible mass-splitting between the components of the multiplet, \emph{i.e}, $m_{R_2} \approx m_{R_2^{5/3}}\approx m_{R_2^{2/3}},$ so as to avoid constraints from electroweak oblique corrections~\cite{Dorsner:2016wpm}.

Rotating into the flavour eigenbasis, we adopt the following convention
\begin{equation}
 \begin{aligned}
 \mathfrak{R}_e \lambda_{eu}\mathfrak{R}_u\mapsto  y^{Seu},\;\;\;\mathfrak{L}_e \lambda_{LQ}\mathfrak{L}_u\mapsto y^{SLQ}, \\
 \mathfrak{L}_e^\dagger \lambda_{Lu}\mathfrak{R}_u\mapsto y^{RLu},\;\;\;
{\mathfrak{R}_e^\dagger}\lambda_{eQ} \mathfrak{L}_u \mapsto y^{ReQ}. \label{mappings}
 \end{aligned}
 \end{equation}
 \noindent Here, $\mathfrak{L}$ and $\mathfrak{R}$ represent the basis mapping between the gauge and flavour eigenstates, and $V = \mathfrak{L}_u^\dagger \mathfrak{L}_d$ is the standard CKM matrix. This is the so-called `up-type' mass-diagonal basis\footnote{ The alternative basis choice would be to select the down-type quark couplings to fix, and allow associated up-type quark couplings to be generated via CKM mixing. This basis encourages ans\"{a}tze that lead to scenarios for $\Delta a_{\mu,e}$ that are ruled-out by significant contribution to $\mu \to e \gamma$~\cite{Crivellin:2018qmi, Dorsner:2020aaz}, so we prefer the up-type basis.} -- a basis choice that we discussed in detail in reference~\cite{Bigaran:2020jil}. This leaves us with the following
\begin{equation}
 \begin{aligned}
 \mathcal{L}^{S_1} \supset y^{SLQ}_{ij} \left[ \overline{e_{L,i}^c} u_{L,j} - V_{jk} \;\overline{\nu_{L,i}^c} d_{L,k} \right]S_1^\dagger\\ + y^{Seu}_{ij} \overline{e_{R,i}^c} u_{R,j} S_1^\dagger+ h.c., \label{lags}
 \end{aligned}
 \end{equation}
   \begin{equation}
 \begin{aligned}
 \mathcal{L}^{R_2} \supset &y^{RLu}_{ij} \left[ \overline{\nu_{L,i}} u_{R,j} R_2^{2/3, \dagger} -\; \overline{e_{L,i}} u_{R,j} R_2^{5/3, \dagger} \right]\\
 &+y^{ReQ}_{ij} \overline{e_{R,i}}\left[ u_{L,j}R_2^{5/3, \dagger} +V_{jk}   d_{L,k}R_2^{2/3, \dagger}\right]\\
 &+ h.c.\label{lagr}
 \end{aligned}
 \end{equation}
Matching these Lagrangians onto equation~\eqref{genericLQ} gives the equivalences
\begin{align}
   &\text{for $S_1$:}\;\; \mathbf{y}^R =\mathbf{y}^{Seu}\;\;\;\;\; \text{and}\; \mathbf{y}^L= \mathbf{y}^{SLQ},\label{coup1}\\
&\text{for $R_2$:}\;\;\mathbf{y}^R=-\mathbf{y}^{RLu}\; \text{and}\;\mathbf{y}^L=\mathbf{y}^{ReQ}.\label{coup2}
\end{align}

We note that only $R_2^{5/3}$ and $S_1$ have couplings of both chiralities with charged-leptons, and that these couplings are to up-type SM quarks. Therefore, dominant contributions to the processes described by Figure~\ref{feyndi} are from up-type quark loops, and for MDMs via the quark-mass enhanced mixed chiral term in equation~\eqref{sigmaL}.

\subsection{\texorpdfstring{$S_1$ and $R_2$ coupling restrictions and the implications for lepton EDMs}{S1 and R2 coupling restrictions and the implications for lepton EDMs}
}
\label{sec:SLQ_sub:EDM}

In reference~\cite{Bigaran:2020jil}, we explored $S_1$ and $R_2$ models to tackle the anomalies in the muon and electron MDMs. There we restricted the LQ coupling texture to represent a dominant single charm-loop contribution to the electron dipole moments, and top-loop to the muon dipole moments
\begin{align}
\mathbf{y}^{L} \sim \begin{pmatrix} 0 &  \colorbox{Gray}{\color{Gray}'}& 0 \\ 
0 &0  & \colorbox{Gray}{\color{Gray}'} \\
 0 & 0 & 0
\end{pmatrix},\;\;\;\;
\mathbf{y}^{R} \sim \begin{pmatrix} 0 &  \colorbox{Gray}{\color{Gray}'}& 0 \\ 
 0 &0  & \colorbox{Gray}{\color{Gray}'} \\
0 & 0 & 0\end{pmatrix}
\label{yukawas2}  \end{align} \noindent where only grey-shaded entries are nonzero, and by convention rows are labelled as charged-lepton, and columns as up-type quark, generations. First-generation quark couplings are avoided as these are known to be highly constrained.  Allowing both anomalies to be generated by the same flavoured up-type quark leads to dangerous contributions to $\mu\to e\gamma$~\cite{Crivellin:2018qmi,Dorsner:2020aaz}, and the alternative coupling texture associated with a \emph{charmphilic} solution to $\Delta a_\mu$ was ruled-out at one-sigma in reference ~\cite{Kowalska:2018ulj}.\footnote{We have performed some estimates of the contributions to $\mu\to e \gamma$ at two-loop level in these models, and find these not to be very constraining on the parameter space given the current upper-bound. Nevertheless, this warrants further careful investigation to determine whether future measurements could provide a sensitive probe for these effects. For an EFT analysis of renormalisation group running effects on lepton dipole moments, see ref.~\cite{Aebischer:2021uvt}.}

We emphasise that in reference~\cite{Bigaran:2020jil} we restricted the coupling entries to be real-valued, in order to avoid constraints from CP-violating processes. Upon conducting a phenomenological study, we found that both models could provide comprehensive solution to the $\Delta a_{e,\mu}$ puzzles so long as the mass of either leptoquark is $\lesssim 65$ TeV.\footnote{In reference~\cite{Bigaran:2020jil}, we did not consider the effects of fine-tuning and leptonic mass corrections which would tighten the upper-bound on LQ mass, as discussed in Section~\ref{sec:LeptonMasses}.} Note also that this analysis was performed before the latest FNAL result for $\Delta a_\mu$, and before the publication of the Rb result for $\Delta a_{e}$. Considering these updated results, we arrive at the constraints on the combinations of couplings shown in Table~\ref{table:Constraint}\footnote{We acknowledge that there is a negative sign error in the approximate coupling constraints derived in equations (30)-(33) of reference~\cite{Bigaran:2020jil} which has been corrected here. It did not impact any other derived results in that paper. }.

\subsubsection{Extending the parameter space to complex couplings}
\label{sec:SLQ_subsub:EDMcomplex}

\begin{table}[t!]
 \begin{center}
  \def\arraystretch{1.8} 
    \begin{tabular}{|c|c|r|} 
    \hline
      Leptoquark & Couplings & Approx. range $\times\left({\hat{m}_\phi^2}\right)$ \\
     \hline \hline
       $S_1$ &  Re$[y^{L,*}_{23} y^{R}_{23}]$ & $(3.1\pm 0.7)\times  10^{-3}$\\
        & Re$[y^{L,*}_{12} y^{R}_{12}]_\text{Cs}$ &$-(4.6\pm 1.9)\times  10^{-3}$\\
        &Re$[y^{L,*}_{12} y^{R}_{12}]_\text{Rb}$ & $(2.5\pm 1.0)\times  10^{-3}$\\
    \hline 
        $R_2$ &  Re$[y^{L,*}_{23} y^{R}_{23}]$ & $-(1.8\pm 0.4)\times  10^{-3}$\\
        & Re$[y^{L,*}_{12} y^{R}_{12}]_\text{Cs}$ &$(4.1\pm 1.7)\times  10^{-3}$\\
        &Re$[y^{L,*}_{12} y^{R}_{12}]_\text{Rb}$ & $-(2.2\pm0.9)\times  10^{-3}$\\  
        \hline \hline
             Leptoquark & Couplings & Upper bound $\times \left(\hat{m}_\phi^2\right)$ \\
     \hline \hline
     $S_1$ &  $|$Im$[y^{L,*}_{23} y^{R}_{23}]|$ & $<1.7$\\
     &  $|$Im$[y^{L,*}_{12} y^{R}_{12}]|$ & $<6.2\times 10^{-9}$\\
         \hline 
        $R_2$ &  $|$Im$[y^{L,*}_{23} y^{R}_{23}]|$ &$<1.1$\\
        &  $|$Im$[y^{L,*}_{12} y^{R}_{12}]|$ & $<5.6\times10^{-9}$\\
        \hline
   \end{tabular}
  \end{center}
    \caption{Approximate one-sigma allowed ranges for the combination of couplings in the chirally-enhanced contribution to leptonic MDMs, and upper bounds for combinations of LQ couplings for contributions to leptonic EDMs. The Cs and Rb fine-structure constant implications for $a_e$ are treated separately.} \label{table:Constraint}
    \end{table}

Naturally, the next step for exploring the implications of these models is to loosen this real-valued assumption on the LQ Yukawa couplings. This opens up a rich parameter space, most notably leading to non-zero contributions to lepton EDMs. The authors of references~\cite{Crivellin:2018qmi,Crivellin:2019mvj} utilise an EFT framework to assess the impact of chirally-enhanced models for the $\Delta a_{e,\mu}$ on the size of lepton EDMs. In this work, we will present a concrete realisation of this link in the context of single scalar LQ models.

Following from equation~\eqref{agamma}, the contribution to the lepton EDMs can be parameterised by
\begin{equation}
\begin{aligned}
    \mathcal{L}_{\ell \gamma}&\supset -\frac{1}{2}id_\ell \overline{\ell} \sigma_{\mu\nu}\gamma_5\ell F^{\mu\nu},\\
       &\supset \frac{1}{2}  e\overline{\ell}  \sigma_{\mu\nu}F^{\mu\nu}\left( \sigma_L^\ell P_L+ \sigma_R^\ell P_R\right)\ell.\label{genericdl}
\end{aligned}
\end{equation}
\noindent Matching coefficients in equation~\eqref{genericdl}, we find that
\begin{align}
    d_\ell = \frac{e}{2i} (\sigma_L^{\ell}-\sigma_L^{\ell*} )= \text{Im}(\sigma_L^{\ell})\; e\,[\text{GeV}^{-1}].
    \label{dedef}
\end{align}
 Note that a consequence of equation~\eqref{sigmaL} is that for pure-real Yukawa couplings, $\text{Im}(\sigma_L^{\ell})$ is exactly zero. The parameter space explored in reference~\cite{Bigaran:2020jil} is therefore a subspace of that explored in this work.
 
In equation~\eqref{dedef}, the squared brackets indicate the units of the result if masses are input in GeV.\footnote{ We also recognise that 
$\text{GeV}^{-1} = 1.98 \times 10^{-14}$ cm, for conversion to appropriate units for EDM measurements.} In Table~\ref{table:Constraint} we provide an overview of the approximate upper-bounds on LQ couplings for satisfying the constraints from $d_\ell$ measurements. Here, and later in this work, we use the following notation for brevity
\begin{align}
    \hat{m}_\phi \equiv \frac{{m}_\phi}{\text{TeV}}.
\end{align}

The results in Table~\ref{table:Constraint} imply a bound on the phase difference between the input couplings relevant for both MDM and EDM of each lepton. We may write

\begin{align}
    y_{ij}^{L*}y^{R}_{ij}=\left|y_{ij}^{L*}y^{R}_{ij}\right|\exp(i \xi_{ij}),
\end{align}
where $\xi_{ij} \in [0, 2\pi)$ is given by
\begin{align}
    \xi_{ij}= {\rm Arg} (y_{ij}^{R})- {\rm Arg}(y_{ij}^{L})
\end{align}
For electrons, assuming that the central value of either MDM result is satisfied, the bound implies that
\begin{align}
    |\tan(\xi_{12} )| \lesssim 10^{-6}, \label{commentTangent}
\end{align}
This is satisfied for angles very close to $\xi_{12}=0$ or $\pi$. In other words, the relative phase between the RH and LH couplings is extremely small because satisfying the EDM constraints severely restricts the size of the imaginary component. The current bound of the muon EDM considered together with the MDM anomaly does not restrict the muon relative phase to the anywhere near the same extent. 

\subsubsection{Counting the free parameters}
\label{sec:SLQ_subsub:EDMfreepar}
The CP-violating phases introduced by extending the parameter space to include complex-valued couplings \emph{a priori} yield four additional parameters for each LQ model. However, by rephasing fields it is straightforward to show that only two of these phases are physical when neutrino masses are neglected. They can be taken to be the relative phases between the exotic RH and LH couplings for the electron-charm and muon-top subsectors. In the limit that each such phase approaches zero, an approximate CP-symmetry is restored to its respective subsector. By this argument, we can say that setting $\xi_{12}=0$ or $\pi$ is technically natural for that subsector considered in isolation, with the overall technical naturalness of the full theory being an open question.

For the remainder of this work we proceed in such a way that we set $\xi_{12}=0$ or $\pi$, and therefore do not generate a contribution to the electron EDM. This is consistent with the preference of current experimental bounds demonstrated by equation~\eqref{commentTangent}. The choice between $0$ or $\pi$ corresponds to scanning over both positive and negative values for the relevant couplings between the electron and charm quark.

\section{Fine-tuning and corrections to the muon and electron masses}
\label{sec:LeptonMasses}
\begin{figure}[t!]
\centering
\includegraphics[scale=1.2]{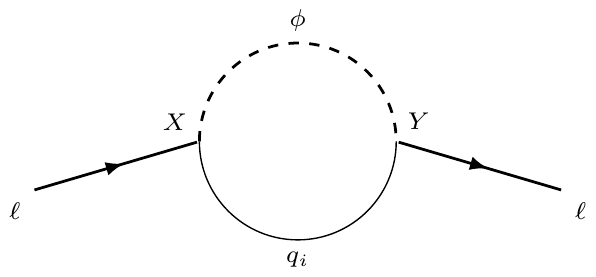}
 \caption{ One-loop correction from scalar LQ to the lepton~($\ell$) propagator. Vertex chirality is labelled $X$ and $Y$ = $\{R,L\}$ for reference in Section~\ref{sec:LeptonMasses}} \label{feyndi2}
    \end{figure}

A sizable, chirally-enhanced, contribution to lepton dipole moments also, consequentially, generates a lepton mass correction\footnote{See reference~\cite{Baker:2021yli} for a discussion of how these class of models could be responsible for lepton (specifically, muon) mass generation.}.  Theoretically, this can be absorbed by adjusting the pole mass, $m_{\ell}^0$.  However, a critique of these simple models is that having this correction greater than the physical lepton mass introduces a fine-tuning problem~\cite{Fuyuto:2018scm,Athron:2021iuf}. In this section, we outline the constraint that this requirement introduces for our models. Note that for simplicity we have switched-off quartic couplings between the Higgs and LQs, but acknowledge that these additional free parameters could extend the viable parameter space for these models -- although, they may also introduce a new suite of fine-tuning problems. The Feynman diagram for this radiative correction is shown in Figure~\ref{feyndi2},

\begin{figure*}[t!]
    \centering
    \includegraphics[width=\textwidth]{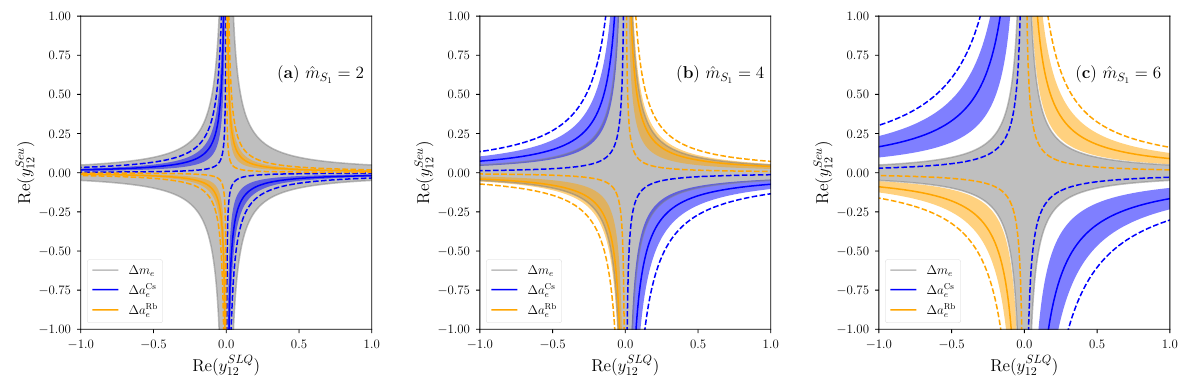}
    \caption{Implications of the constraint on the electron mass correction on $S_1$ LQ solutions to the electron MDM, for (a) $m_{S_1}= 2$ TeV , (b) $m_{S_1}= 4$ TeV and (c) $m_{S_1}= 6$ TeV. These illustrations of parameter space assume that the imaginary components of the electron LQ couplings are negligible. The shaded grey region, common in all plots, shows the allowed points by restriction on the electron mass correction in equation~\eqref{masscon}. For the Cs and Rb MDM results, solid lines show the central values, shaded regions are the one sigma region, and the two sigma region is denoted by a dashed line.}
    \label{fig:ElectronMassS1}
\end{figure*}
Referring again to Figure~\ref{feyndi2}, the contribution of this correction to a charged-lepton  mass, $m_\ell$, is defined via a one-loop self energy, $\Sigma_\ell$. For the full calculation of these corrections, we refer to reference~\cite{Crivellin:2020mjs}, but quote the relevant results here. Up to linear order in the self-energy expansion, the lepton mass correction is
\begin{align}
    \delta m_\ell&=m_{\ell}^0\left( \frac{1}{2}\Sigma_\ell^{LL}+\frac{1}{2}\Sigma_\ell^{RR}\right)+\Sigma_\ell^{LR},
   \end{align}
   where $\Sigma_\ell^{XY}$= $\Sigma_\ell^{XY}(p^2=0)$, \emph{i.e} we take the momentum independent contributions of the self-energies to generate a physical mass correction. Here $X$ and $Y$ refer to the chirality of the labelled vertices in Figure~\ref{feyndi2}.
   
   The general expressions for these self-energies evaluated at energy scale $\mu$ are given by
   {\small
   \begin{align}
       \Sigma_{\ell}^{LR}& = -\sum_q \frac{m_{q} N_c}{16 \pi^2} \mathcal{I}_0 \left(\frac{\mu^2}{m_\phi^2}, \frac{m_q^2}{m_\phi^2}\right)y^{R*}_{\ell q} y^{L}_{\ell q},\\
       \Sigma_{\ell}^{LL}+\Sigma_{\ell}^{RR}&= -\frac{ N_c}{32 \pi^2}\sum_q \mathcal{I}_1 \left(\frac{\mu^2}{m_\phi^2}, \frac{m_q^2}{m_\phi^2}\right)\left[|y^{L}_{\ell q}|^2+|y^{R}_{\ell q}|^2\right].
   \end{align}}
 where the loop functions are given by
 \begin{align} 
 \mathcal{I}_0 (x,y)& = \frac{1}{\epsilon} + 1+ \log(x) +y\log(y),\\
  \mathcal{I}_1 (x,y)& = \frac{1}{\epsilon} + \frac{1}{2}+ \log(x)-y,
 \end{align}
 and the couplings are as defined in equations~\eqref{coup1} and~\eqref{coup2}. Evaluating the masses at the LQ scale $\mu=m_\phi$, and taking only the finite contributions to $m_\ell$, we presume to cancel the self-energy divergences via $\overline{MS}$ as per reference~\cite{Crivellin:2020mjs}. This yields the following:
 \begin{align}
     \mathcal{I}_0 (1,m_q^2/m_\phi^2) &\mapsto 1+ m_q^2/m_\phi^2\log(m_q^2/m_\phi^2)\\
     \mathcal{I}_0 (1,m_q^2/m_\phi^2) &\mapsto 1- m_q^2/m_\phi^2
 \end{align}
Assuming dominance of the mixed-chiral contribution $\Sigma_{\ell}^{LR}$ and we require for both leptons that
 \begin{align}
    \left|\frac{\delta m_\ell}{m_{\ell}^0}\right|&=\left|\frac{\Sigma_\ell^{LR}}{m_{\ell}^0}\right|\lesssim 1, \label{masscon}
   \end{align}
which is to say, it can account for the lepton mass but should not require significant, tree-level, fine-tuning. Therefore, now noting that in our model only a single quark species runs in the loop at a time, we can drop the summation over $q$ and we arrive at
\begin{align}
     |\delta m_\ell|& \simeq  \frac{m_{q} N_c}{16 \pi^2} |y^{R*}_{\ell q} y^{L}_{\ell q}|\left[1+\frac{2m_q^2}{m_\phi^2}\log\left(\frac{m_q}{m_\phi}\right)\right]. \label{deltaml}
\end{align}
   
We take the pole masses of the two leptons at the TeV scale to be  $m_\mu^0=0.10468806$ GeV and $m_e^0=4.959016 \times 10^{-4}$ GeV~\cite{Xing:2007fb}. We will explore the implications of the correction on the electron and muon mass, and their interplay with dipole moments, below.

\subsection{Preliminary comment on the considered LQ masses}
\label{sec:LeptonMasses_sub:prelim}

Presently the scalar LQ mass, $m_\phi$, is most strongly constrained using LHC searches for the decay of pairs of scalar LQs with couplings predominantly to first-generation leptons~\cite{Aad:2020iuy}, \begin{equation}{\hat{m}_\phi}> 1.8\;\; \text{at}\; 95\% \text{CL}.\label{masslim}\end{equation} 
Given that we allow such couplings in the models detailed in this work, and these couplings can take order-one values, we adopt equation~\eqref{masslim} as a conservative lower-bound on the LQ mass. 

For the remainder of section, we selected three benchmark LQ masses for specific study
\begin{align}
    {\hat{m}_\phi} \in \{2, 4, 6\}.
\end{align}
which we will see span the allowed regions of parameter space, and extend to those forbidden by the mass corrections and not preferred by the present dipole measurements. This is discussed further in forthcoming sections.

\subsection{Electron MDM implications for the electron mass}
\label{sec:LeptonMasses_sub:electron}

  We elaborate further here on the constraints that the electron mass, in particular, places on the viability of these models for explaining the anomalies in the electron AMM. For this study, we assume that the imaginary component for both the electron RH and LH couplings are negligible, \emph{i.e.} $\xi_{12} =$ 0 or $\pi$.
       
Combining the electron result with the central values for MDM results, we can derive a conservative bound
\begin{align}
   \hat{m}_\phi\lesssim 3.12\;. \label{electronbound}
\end{align}

Figure~\ref{fig:ElectronMassS1} shows a cross section of $S_1$ parameter space. We find that the upper bound on the couplings from the correction to $m_e$ is largely LQ mass independent for the considered masses. It is evident from Figure~\ref{fig:ElectronMassS1}(a), that for smaller masses this electron mass correction is small for solutions to both MDM results. For a mass of 4 TeV, this model can no longer reach the central value for the Cs anomaly, whereas there is sizeable parameter space remaining to explain the Rb result. At a mass of 6 TeV, it appears that for the region of parameter space illustrated, the $S_1$ leptoquark cannot accommodate the Rb or Cs MDM result, within one sigma. 

Note that the $R_2$ leptoquark was found to demonstrate similar behaviour in the $y^L_{12}- y^R_{12}$ plane, but with reflection over the Re$(y^L_{12})$ and Re$(y^L_{12})$ axes, illustrating the preference for opposite-signed coupling combinations.

\begin{figure*}[t!]
    \centering
    \includegraphics[width=\textwidth]{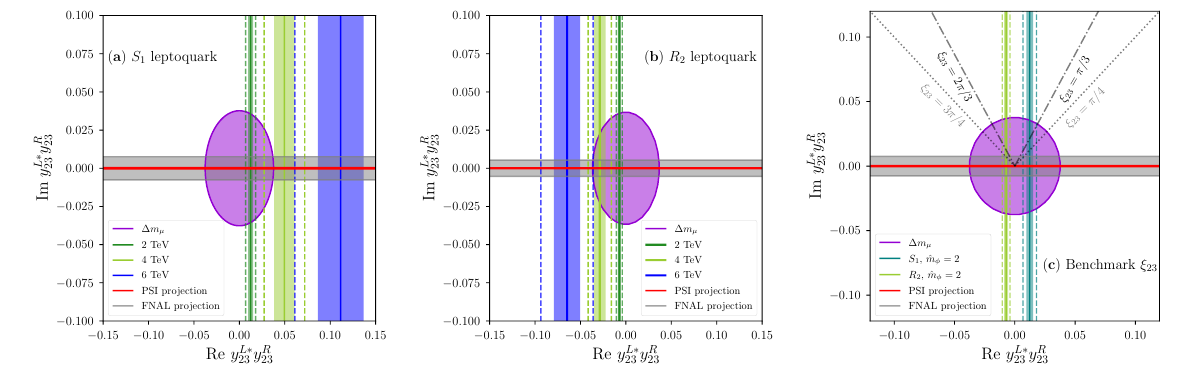}
   \caption{Implications of the constraint on the muon mass correction on $S_1$ and $R_2$ LQ solutions to the muon dipole moments. Subfigure (a)~($S_1$) and  subfigure (b)~($R_2$) demonstrate the interplay of constraints for benchmark masses $\hat{m}_\phi =2, 4, 6$. For each mass, solid lines show the central values, shaded regions are the one sigma region, and the two sigma region is denoted by a dashed line. Horizontally bound regions show the projected reach of PSI and FNAL, such that the coloured regions represent parameter space that would generate an EDM beyond their reach. In subfigure (c) we depict benchmark phases $\xi_{23}$ for a mass of $\hat{m}_\phi =2$, demonstrate the overlap with viable muon MDM solutions, for each LQ. Contours representing the trivial phases $\xi_{23}=0(\pi)$ are not explicitly shown, but would lie along the positive (negative) x axis.}
    \label{fig:MuonMass}
\end{figure*}     
\subsection{Interplay of corrections to the muon mass and muon dipole moments}
\label{sec:LeptonMasses_sub:muon}

We first consider the muon EDM and the relationship between dipole contributions and the muon mass correction, without considering the muon AMM. Following from equation~\eqref{deltaml}, we can derive a bound on the muon coupling magnitudes,
\begin{align}
    |y_{23}^L y_{23}^R|= \sqrt{{\rm Im}\left(y_{23}^{L,*}y_{23}^R\right)^2+{\rm Re}\left(y_{23}^{L,*} y_{23}^R\right)^2}\lesssim 0.04,
\end{align}
which is roughly independent of mass for the considered mass-range. Now, we can arrive at a conservative upper bound on the imaginary component by letting ${\rm Re}\left(y_{23}^L y_{23}^R\right) \to 0$,
\begin{align}
|{\rm Im}\left(y_{23}^{L,*}y_{23}^R\right)|\lesssim 0.04 \label{boundonImMuon}
\end{align}
Following from equation~\eqref{dedef}, we can derive a prediction for the upper bound for the $S_1$ and $R_2$ muon EDM, and therefore a measure of at what point future experiments will be sensitive to these models. This gives the following
\begin{align}|d_\mu|_{S_1} 
 \lesssim \left\{
\begin{array}{*{2}{c}}
2.77,& \hat{m}_\phi = 2 \\
0.96,& \hat{m}_\phi = 4 \\
0.50,& \hat{m}_\phi = 6 \\
\end{array} \right\}\times 10^{-22}\; e\;\text{cm};
\end{align}
\begin{align}|d_\mu|_{R_2} 
 \lesssim \left\{
\begin{array}{*{2}{c}}
3.91,& \hat{m}_\phi = 2 \\
1.25,& \hat{m}_\phi = 4 \\
0.62,& \hat{m}_\phi = 6 \\
\end{array} \right\}\times 10^{-22}\; e\;\text{cm}.
\end{align}

The Paul Scherrer Institut~(PSI) muEDM experiment is projected to have a sensitivity within the ballpark of $5\times 10^{-23}$ $e$~cm~\cite{Kirch:2020lbo,Adelmann:2021udj}. Similarly, the FNAL Muon g$-$2 experiment aims to probe muon EDMs of order $10^{-21}e$~cm~\cite{Chislett:2016jau}. Therefore, muon EDMs generated by these models are well within the reach of the PSI experiment, and maybe even by FNAL depending on the value of $\xi_{23}$. Projected reach of both experiments can be seen in Figure~\ref{fig:MuonMass} as horizontally bound regions, such that the coloured regions represent the remaining parameter space that would generate an EDM beyond their future sensitivity.

We now analyse the implications for the muon MDM from the fine-tuning issue raised by muon mass corrections. This interplay can be seen explicitly in Figure~\ref{fig:MuonMass}. We cannot immediately assume the relative suppression of the imaginary component of the coupling component to the real, as we did for the electron as a result of current experimental bounds. We can, however, recognise that a similar logic to that which goes into deriving equation~\eqref{boundonImMuon} can be applied here, to achieve the bound (agnostic to masses within the considered range),
\begin{align}
|{\rm Re}\left(y_{23}^{L,*}y_{23}^R\right)|\lesssim 0.04, \label{boundonReMuon}
\end{align}
which is required to ensure that the muon mass correction is not too large. Combining this with the bounds quoted in Table~\ref{table:Constraint} we can arrive at the following mass constraint for these models for explaining the muon AMM: for the $S_1$ LQ,
\begin{align}\hat{m}_{S_1} 
 \lesssim \left\{
\begin{array}{*{2}{c}}
4.08,& \text{at}\; \text{CV}\\
4.71,& \text{at}\;1\;\sigma \\
4.85,& \text{at}\;2\;\sigma \\
\end{array} \right\},
\end{align}
and for the $R_2$ leptoquark,
\begin{align}\hat{m}_{R_2} 
 \lesssim \left\{
\begin{array}{*{2}{c}}
3.59,& \text{at}\; \text{CV}\\
5.36,& \text{at}\;1\;\sigma \\
6.32,& \text{at}\;2\;\sigma \\
\end{array} \right\}.
\end{align}
where CV denotes the present central value for the muon MDM anomaly, followed by the one and two sigma limits. These upper mass limits are generally applicable to any $S_1$ or $R_2$ LQ model whose contribution to the muon MDM is generated by a top-containing loop. Note that for either of our models, these mass constraints are less stringent than that derived from the electron mass correction in equation~\eqref{electronbound}.

The interplay of above constraints can be seen in Figure~\ref{fig:MuonMass}. In Figure~\ref{fig:MuonMass}(c) we focus on a LQ mass of $\hat{m}_\phi =2$ and show contours representing benchmark values of the phase parameter \begin{align}
 \label{phase1} \xi_{23}=  \{0,\frac{\pi}{4}, \frac{\pi}{3}\} \;\; \text{for}\; S_1\\
  \xi_{23}=\{ \pi, \frac{2\pi}{3}, \frac{3\pi}{4}\}\;\; \text{for}\;R_2,\label{phase2}
\end{align} consistent with the sign preference for each model, as detailed in Table~\ref{table:Constraint}. Contours representing the trivial phases $\xi_{23}=0(\pi)$ are not explicitly shown, but would lie along the positive (negative) x axis. For the non-trivial phases shown, an EDM generated by either of the LQ models satisfying the muon MDM central value, could feasibly be probed by either PSI or the FNAL experiment.

\section{\label{sec:Additional Constraints} Additional constraints}
\label{sec:Additional}

Many of the additional constraints considered in this work overlap with those considered in reference~\cite{Bigaran:2020jil}, particularly in the kaon sector, for high $p_T$ tails, and for leptonic $Z$ decays. Here we provide an overview of these and additional constraints, as well as indicating the importance of some future probes of our models. A summary of present experimental values for the pertinent constraints can be found in Table~\ref{table:const}.

\subsection{Contributions to composite EDM measurements}
\label{sec:Additional_sub:compEDM}
The contributions of the LQ couplings to QCD effective interactions at $\mu=m_\phi$  are very small, as $\alpha(m_\phi)$ is small. Nevertheless, the RGE evolution of the generated four-fermion interactions permit mixing into these contributions as they evolve to the low-scale. This makes the contributions from the effective quark and gluon (chromo) EDMs important to consider, as these could lead to nonzero predictions for nucleon EDMs, and those of paramagnetic, polar molecular systems~\cite{Fuyuto:2018scm}. 

 Thus far, EDM searches in general have null results, yet they yield stringent upper bounds. For muon-top couplings of LQs, Dekens \emph{et al}~\cite{Dekens:2018bci} find that the muon EDM provides stronger constraints than the composite measurements. Similarly, for the electron-charm couplings, the $d_e$ constraint is most constraining. 
 
 With improved modelling of composite structures these constraints may become tighter, but for now and in the foreseeable future our model couplings are most strongly constrained by the charged-lepton EDMs. For this reason, we do not discuss these additional EDM contributions further. A detailed discussion on how the $R_2$ and $S_1$ leptoquarks can generate fundamental and composite EDMs is provided in reference~\cite{Dekens:2018bci}.
 
\begin{table}[h!]
 \begin{center}
  \def\arraystretch{2} 
    \begin{tabular}{|c|c|c|} 
    \hline
      \textsc{Process}  & \textsc{Observable}&\textsc{Constraint} \\
     \hline \hline
       $Z \to \ell_i \ell_j$ &  
          $ g_A^e/g_A^{e,\text{SM}}$ &$0.999681 \pm 0.000698227 $ \\
          &  
          $ g_A^{\mu}/g_A^{\mu,\text{SM}}$ &$0.99986 \pm 0.00107726 $ \\
        \hline 
       $Z \to \nu \nu$              &  $N_\nu^\text{Eff.}$  & $2.9840(82)$\\
    \hline
    $K^+\to \pi^+ \nu \nu$ & Br & $(1.7\pm 1.1) \times 10^{-10}$ \\ 
        $K^0_L\to \pi^0 \nu \nu$ & Br & $<2.6 \times 10^{-8}$ \\ 
      $K^0_L\to e^+ e^-$ & Br & $(9^{+6}_{-4}) \times 10^{-12}$  \\ 
      $K^0_L\to \mu^+ \mu^-$ & Br &  $( 6.84 \pm 0.11 ) \times 10^{-9} $ \\
       $K^0_L\to \mu^+ e^-$ & Br &   $< 4.7 \times  10^{-12} $\\
       \hline 
       $K_0$-$\overline{K_0}$ mixing & $|\epsilon_K|$ & $(2.228 \pm 0.011) \times 10^{-3}$ \\
        & $\Delta M_K$ & $(3.484 \pm 0.006) \times 10^{-12}$ \\
    \hline
    $pp\to \ell\ell$ &$|y^{Seu}_{12}|$& $< 0.648~\hat{m}_\phi$ \\
   \cite{Greljo:2017vvb}&$|y^{RLu}_{12}|$& $<0.524~\hat{m}_\phi$\\
        \hline
   \end{tabular}
  \end{center}
    \caption{Processes most constraining on this model. Values quoted without citation are from PDG~\cite{PDG}.  Constraints from $pp\to \ell\ell$ are derived from Table 1 of reference~\cite{Greljo:2017vvb}. } \label{table:const}
    \end{table}

\subsection{\texorpdfstring{Leptonic $Z$ and Higgs decays}{Leptonic Z and Higgs decays}}
\label{sec:Additional_sub:ZandH}

Given the chirally-enhanced nature of contributions to the lepton effective dipole vertex, correlations with other observables sensitive to electroweak symmetry breaking are to be expected~\cite{Crivellin:2021rbq}. In particular, $Z$-boson decays to dilepton final states are expected to constrain the same set of LQ couplings that are sensitive to lepton dipole moments. Also, given the sizable corrections to the lepton masses established above, we would expect BSM effects to also be present in the dilepton decays of the higgs, in particular $h\to \mu\mu$. We discuss these two sets of constraints below.

\subsubsection{Z effective couplings}
\label{sec:Additional_subsub:Zeff}
 The couplings $g_{f_{L(R)}}$ are the effective LH and RH couplings of the Z boson to fermions, $f$. The effective Lagrangian for describing the BSM contributions to these interactions is given by:
\begin{equation}
    \mathcal{L}^Z_\text{eff} = \frac{g}{\cos(\theta_W)} \sum_{i,j} \overline{f_i} \gamma^\mu\left[ g^{ij}_{f_L}P_L+ g^{ij}_{f_R}P_R \right]f_j Z_\mu,
\end{equation}
where $\theta_W$ is the weak-mixing angle. For this study, we use constraints from corrections to effective vector and axial-vector couplings about their central values, where 
\begin{align}
   g_{f_{V(A)}}^{ ij}= g_{f_L}^{ij}\pm  g_{f_R}^{ij}.
\end{align}
For $i=j$, we relabel the coupling $g^{ii}_{f_{A(V)}}\equiv g^{f_i}_{A(V)}$
We follow the calculation of effective $Z$ couplings to charged-leptons $\ell$, and associated observables, from reference~\cite{Arnan:2019olv}. For TeV scale LQs, the top-mass enhancement of the contribution to the effective coupling to the muons gives
{\small
\begin{align}
    \delta g^\mu_{A(V)} \simeq -\frac{N_{c}}{32\pi^2}\frac{m_t^2}{m_\phi^2}\left(1+\log \frac{m_t^2}{m_\phi^2} \right) \left(|y^{R}_{23}|^2 \pm|y^{L}_{23}|^2  \right)
\end{align}}

Moreover, the axial vector couplings are the most tightly constrained, so we focus on these for constraining these models. These constraints are quoted in Table~\ref{table:const}.

Complementary constraints on the LQ interaction with neutrinos can be inferred from the effective number of light neutrino species, $N_\nu^\text{Eff}$, which parameterises the decay rate of the $Z$ to invisible final states ($Z\to \nu\bar{\nu}$). These are particularly important for the $R_2$ leptoquark, where the neutrinos also couple to up-type quarks and these couplings are not subject to CKM suppression. Again, the top-quark contribution is dominant and following reference~\cite{Arnan:2019olv}, the $R_2$ contribution is calculated as
\begin{align}
    N_\nu^\text{Eff} & \simeq 3 -N_{c}\frac{|y^{RLu}_{23}|^2 }{8\pi^2}\frac{m_t^2}{m_\phi^2}\left(1+\log \frac{m_t^2}{m_\phi^2} \right).
\end{align}
For $S_1$, neutrinos only couple to down-type quarks, therefore, no top-mass enhancement, and also these couplings are generated via the CKM. Therefore, for this LQ, $Z\to \nu\nu$ does not provide a competitive constraint.

\subsubsection{Higgs dilepton decays}
\label{sec:Additional_subsub:Zdecays}
The LQ contribution to Higgs decay to the dilepton final state is strongly correlated to the aforementioned lepton mass corrections, and to the electric and magnetic dipole moments.\footnote{Correlations in the context of the EFT between $h\to\mu\mu$ and $\Delta a_\mu$ were considered in references~\cite{Crivellin:2020mjs,Fajfer:2021cxa}, and $h\to\tau\tau$ and $\Delta a_\tau$ in reference~\cite{Feruglio:2018fxo}.} For years LHC measurements of Higgs decays to two charged-leptons were limited to the those of the third generation. Recently, the first evidence for decays $h\to \mu^+\mu^-$ from the ATLAS and CMS collaborations\cite{CMS:2020xwi,ATLAS:2020fzp} indicate that there is an enhancement of the signal strength over the SM prediction, although there is considerable uncertainty. 

Following the derivation in reference~\cite{Crivellin:2020tsz}, we find the precise approximation for the contribution of the $S_1$ leptoquark from a loop containing the quark $q$
{\small
\begin{align}
    \frac{\mathrm{Br}(h\to \ell^+\ell^-)}{\mathrm{Br}(h\to \ell^+\ell^-)_\mathrm{SM}} \approx
    \left|1+ \frac{m_q^3}{8m_\ell} \frac{N_c}{8\pi^2} \frac{y^{R*}_{\ell q} y^{L}_{\ell q}}{m_\phi^2} \mathcal{J}\left(\frac{m_h^2}{m_\phi^2}, \frac{m_q^2}{m_\phi^2}\right) \right|,\nonumber\\
    \;\;\approx 1+ \frac{m_q^3}{8m_\ell} \frac{N_c}{8\pi^2} \frac{\text{Re}(y^{L*}_{\ell q} y^{R}_{\ell q})}{m_\phi^2} \mathcal{J}\left(\frac{m_h^2}{m_\phi^2}, \frac{m_q^2}{m_\phi^2}\right),\label{hll}
\end{align}}where the second line follows from assuming that the magnitude of the LQ correction is small. The form of equation~\eqref{hll} is identical for $R_2$ except that we switch $y^{L}_{\ell q}\leftrightarrow y^{R}_{\ell q}$.\footnote{Reference~\cite{Crivellin:2020tsz} adopts the labelling of the chirality of the interaction by that of the lepton in the LQ interaction, hence the relabelling required.} The loop function here is given by:
\begin{align}
 \mathcal{J}(x, y)= 2 (x - 4) \log(y) - 8 +\frac{13}{3}x.
\end{align}
For the dimuon final state, LQ couplings in equation~\eqref{hll}, Re($y^{R*}_{23} y^{L}_{23}$), also appears in, and is constrained by, the expression for $\Delta a_\mu$. Moreover, the sign of this coupling combination required for the $R_2$ and $S_1$ LQ to generated the measured result for the muon $\Delta a_\mu$ is opposite. This means that the observable $\mathrm{Br}(h\to \mu^+\mu^-)$ provides a probe for distinguishing the effects of these two scalar LQs. Following Table~\ref{table:Constraint}, observing an enhancement over the SM would be consistent with an $S_1$ model for $\Delta a_\mu$, and a suppression would be consistent with that of $R_2$.

We will not discuss this notion further here as the current measurements are not yet sensitive enough to provide a definitive test, but note that this is discussed in detail in references~\cite{Crivellin:2020tsz,Fajfer:2021cxa}. 

We may ponder whether that same probing capacity would be true of $\mathrm{Br}(h\to e^+e^-)$. This is such an extraordinarily rare process in the SM that any measurement of this decay would signal BSM physics~\cite{ATLAS:2019old}. If this process were to be measured at high precision in future experiments, these two scalar LQ models of $\Delta a_e$ \emph{could} be distinguished, but only once the sign conflict of this anomaly is resolved.

\subsection{Constraints from the kaon sector}
\label{sec:Additional_sub:kaon}

 Interactions involving strange-mesons are generated by these LQ couplings via $SU(2)_L$ invariance of the associated interaction, explicitly manifest in equations~\eqref{lags} and \eqref{lagr}. This also means that the constraints here are most directly relevant for refining the parameter space for LH LQ couplings. The implications for scalar LQ models from constraints in the kaon sector have been studied in references~\cite{Kumar:2016omp,Bobeth:2017ecx, Mandal:2019gff}.

 Following the derivations in reference~\cite{Mandal:2019gff}, we re-derive the relevant constraints for each LQ model. As $S_1$ is $|F|=2$ and $R_2$ is $|F|=0$, this makes identifying the strongest constraints heavily model-dependent. We find that the $R_2$ parameter space is much more tightly constrained by the kaon sector than that of $S_1$. 
 
 Additionally, because of the CKM dependence of couplings between the LQs and down-type quarks, we find that these constraints do not tightly constrain parameter phases. This follows from the discussion in Section~\ref{sec:SLQ_subsub:EDMfreepar}, where we have argued that the only physical phases can be phrased as the difference between the LH and RH couplings' phase. Physical observables should be independent of the basis choice, and we can choose a basis in which the phases can all lie in the RH LQ couplings. Thus, we do not expect phase constraints to arise from considering the kaon sector, as a consequence of our coupling ansatz.
 
\begin{figure*}[t!]
   \centering
   \includegraphics[width=\textwidth]{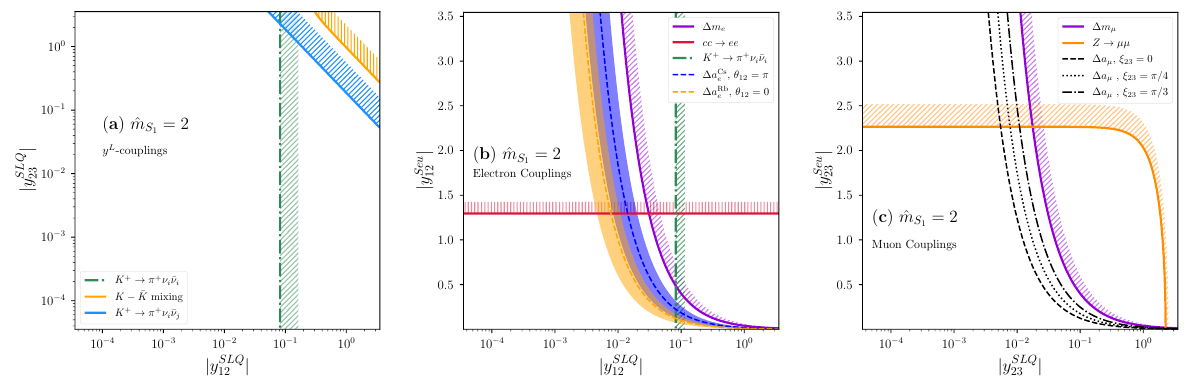}
   \caption{An overview of the constraints on the magnitude of the $S_1$ LQ couplings for a benchmark mass of 2 TeV. The plots show (a) the $y^L$ coupling plane, (b) electron couplings and (c) muon couplings. The hashed regions indicate that the region is ruled out in the direction of the hashing. The $y^L$ couplings are subject to stronger bounds, so their axes are logarithmic while the RH couplings are shown on linear axes. For the electron couplings, the shaded region around the indicated MDM lines show the one-sigma margin of error on the experimental values, from equations~\eqref{g-2eCs} and~\eqref{g-2eRb}. For the muon couplings, the central value is indicated for a number of benchmark angles $\xi_{23}$.}
   \label{fig:sectorS1}
\end{figure*}
 
 \subsubsection{\texorpdfstring{$K^+ \to \pi^+ \nu\overline{\nu}$}{Kpnunu}}
 \label{sec:Additional_subsub:Kpinunu}
Contributions to the theoretically clean rare process Br$(K^+ \to \pi^+ \nu\overline{\nu})$ are found to give competitive constraints for the $S_1$ model. As the effective interaction here is purely vector in nature, it does not run in QCD. Taking the leading-order contribution and following the derivation in reference~\cite{Mandal:2019gff}, we obtain the following bound from the lepton-flavour conserving contribution:\begin{align}
|y_{12}^{SLQ}| < 0.041\; {\hat{m}_\phi},
\end{align}
\noindent
and for the flavour-violating contribution:
\begin{align}
    |y_{12}^{SLQ} y_{23}^{SLQ}|< 0.0467\;{\hat{m}_\phi}^2.
\end{align}

We find constraints on these couplings are stronger for this process in contrast with the alternative decay $K^0_L \to \pi^0 \nu\overline{\nu}$. We also note that given that we permit lepton flavour violating processes, the Grossman-Nir~(GN) bound does not directly apply to this model \cite{Grossman:1997sk,He:2018uey}.The final state contributions that violate lepton flavour can generate CP-conserving contributions to $K^0_L \to \pi^0 \nu\overline{\nu}$, otherwise negligible in the SM. This makes future measurements of this process very likely to generate competitive constraints for the relevant LQ couplings.
Nevertheless, given the relatively low statistics for the current KOTO~\cite{KOTO:2020prk} and NA62~\cite{NA62:2021zjw}, we refrain from discussing the GN bound further here.

\subsubsection{\texorpdfstring{$K^0_{L} \to \ell \ell'$}{K0L to ell ell}}
\label{sec:Additional_subsub:Kll}
These decays are forbidden at tree-level in the SM, and therefore reference~\cite{Mandal:2019gff} conservatively assumes that the LQ contribution saturates the upper limit. Conversely, for the $R_2$ model the strongest constraint from the kaon sector comes from tree-level leptoquark exchange to the helicity suppressed $K^0_L \to e^+ e^-$ transition.  Noting the vector nature of the effective contribution to this process, and following again from reference~\cite{Mandal:2019gff}, we derive an upper bound at $90\%$ confidence on the associated coupling: \begin{align}
    |y^{ReQ}_{12}|\lesssim 0.0956\; {\hat{m}_\phi}.\label{Kee}
\end{align}
Similarly for the muon coupling, the significantly less-suppressed process $K^0_L \to \mu^+ \mu^-$ yields: \begin{align}
    |y^{ReQ}_{23}|\lesssim 0.367 \;{\hat{m}_\phi},\label{Kmm}
\end{align}
and for the lepton-flavour violating decay $K^0_L\to e^\pm \mu^\mp$ yields:\begin{align}
    |y_{23}^{ReQ}y_{12}^{ReQ}| < 3.78\; {\hat{m}_\phi}^2\times 10^{-4} .
\end{align}
For each of these contributions, the couplings involved are between down-type and charged-leptons, generated via CKM mixing. This means that, due to the refined coupling texture, the constraints enter for the magnitudes of these couplings rather than having phase-dependence. These combinations of constraints carve out viable parameter-space for these models in the magnitude ~$|y_{23}^{ReQ}| -|y_{12}^{ReQ}|$ plane.

\subsubsection{\texorpdfstring{Kaon mixing: $|\epsilon_K|$}{Kaon mixing: epsK}}
\label{sec:Additional_subsub:Kmixing}
An additional constraint on these models arises from considering neutral kaon mixing. Because of the restricted coupling structure, dependence on the phases of the input couplings drop out, however the relative sizes are constrained by the parameter $\epsilon_K$, such that
\begin{align}
     |y_{23}^{L}|^2(|y_{12}^{L}|^2&-0.0016|y_{23}^{L}|^2)\nonumber  \\
     &\in [-0.077, 0.232]\times\; {\hat{m}_\phi}^{2}.
\end{align}
The above relation holds for both $S_1$ and $R_2$, derived following reference~\cite{Mandal:2019gff}. In both instances, this constraint is found to be stronger for this model than the complementary constraint from neutral kaon mixing $\Delta M_K$. For the considered mass range, the effects on kaon mixing are a less stringent constraint than those from other kaon sector observables discussed above. This is based on the current constraints listed in Table~\ref{table:const}.

\subsection{\texorpdfstring{High $p_T$ leptonic constraints}{High pT leptonic constraints}}
\label{sec:Additional_subsub:pT}
We briefly comment here on the constraints from contributions to $pp\to \ell \ell$ (dilepton production) and $pp\to \ell \nu$ (monolepton production) via tree-level t-channel processes. These can be probed directly in the high-$p_T$ tails of Drell-Yan processes at the LHC~\cite{Aaboud:2017buh}. For these processes, numerical analyses find minimal interference between NP effective operator contributions and \emph{a priori} several NP operators can be simultaneously constrained~\cite{Greljo:2017vvb}.

In reference~\cite{Bigaran:2020jil} we presented an outline of the impact of these constraints on the magnitudes of couplings in both the $R_2$ and $S_1$ model. As these constraints also apply to the parameter space explored in this paper, we refrain from discussing the details of these calculations here but direct the reader there for further detail. 
Only RH coupling constraints as consequence of the dilepton high-$p_T$ study are listed in Table~\ref{table:const}. The LH couplings are more tightly constrained by the meson decays discussed earlier. Of the constraints from charm monolepton decays derived from reference~\cite{Fuentes-Martin:2020lea}, none were found to be more constraining than the requirement for the LQ couplings to generate the MDM experimental values.

Note that with increased LHC luminosity, the allowed regions for the effective interactions manifest in high-$p_T$ leptonic tails are forecast to shrink significantly~\cite{Fuentes-Martin:2020lea, Greljo:2017vvb}, further restricting parameter space for both models. However, we limit our consideration here to present high-$p_T$ constraints.

\subsection{Summary of additional constraints}
\label{sec:Additional_sub:summary}
We select a benchmark mass of $\hat{m}_\phi=2$ to provide demonstrative cross sections of these models' parameter space. Our aim here is to show the interplay between different additional constraints discussed in this section, together with the constraints from lepton mass correction and satisfying the MDM experimental values.

\begin{figure*}[t!]
   \centering
   \includegraphics[width=\linewidth]{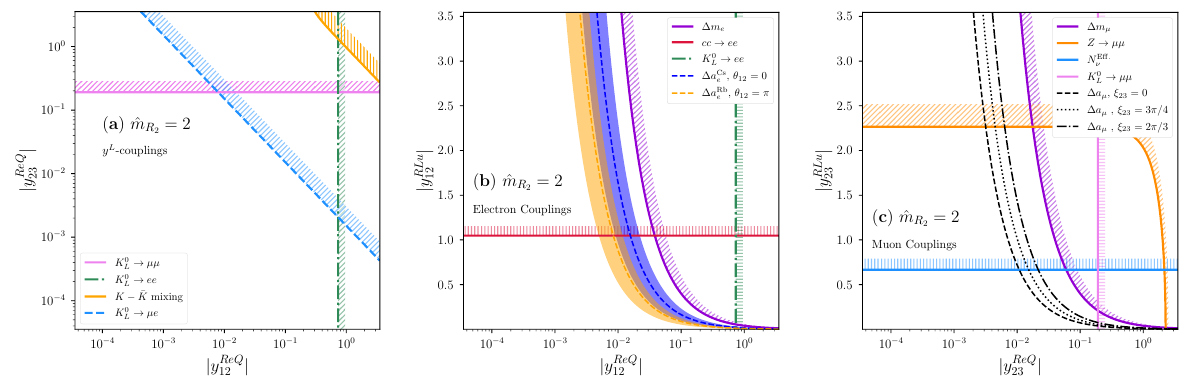}
   \caption{An overview of the constraints on the magnitude of the $R_2$ LQ couplings for a benchmark mass of 2 TeV. The plots show (a) the $y^L$ coupling plane, (b) electron couplings and (c) muon couplings. The hashed regions indicate that the region is ruled out in the direction of the hashing. The $y^L$ couplings are subject to stronger bounds, so their axes are logarithmic while the RH couplings are shown on a linear axis. For the electron couplings, the shaded region around the indicated MDM lines show the one-sigma region about the central value on the experimental values, from equations~\eqref{g-2eCs} and ~\eqref{g-2eRb}. For the muon couplings, the central value is indicated for a number of benchmark angles $\xi_{23}$.}
   \label{fig:sectorR2}
\end{figure*}

Firstly, Figures~\ref{fig:sectorS1}(a) and~\ref{fig:sectorR2}(a) show the plane of $y^L$ coupling magnitudes, $|y^L_{23}|$ and $|y^L_{12}|$. The constraints appearing here originate from the kaon sector. We see that the $R_2$ leptoquark parameter space is much more strongly constrained than that of $S_1$, as discussed in Section~\ref{sec:Additional_sub:kaon}.

In Figures~\ref{fig:sectorS1} and~\ref{fig:sectorR2} we show the $|y^L_{ij}|$ and $|y^R_{ij}|$ parameter space separately for electron and muon sectors, in subfigures (b) and (c) respectively. The hashed contours show parameter space that is ruled-out by an array of constraints detailed in this section. In the central plots,  we illustrate regions that satisfy the electron MDM constraints for Cs and Rb, to within one-sigma. For the muon sector in Figures~\ref{fig:sectorS1}(c) and ~\ref{fig:sectorR2}(c), we show contours representing the benchmark values of $\xi_{23}$, consistent with the sign preference for each model -- as per equations~\eqref{phase1} and \eqref{phase2}.

One way to interpret the results in Figures~\ref{fig:sectorS1} and~\ref{fig:sectorR2} is as follows:
\begin{enumerate}
    \item Identify a point on subfigure (c) that is consistent with the muon MDM, and within experimental constraints. Note the LH coupling at this point;
    \item Move to subfigure (a) and use the LH muon coupling to find a LH electron coupling consistent with constraints; 
    \item Use the electron coupling to check subfigure (b) for compatibility with either electron MDM result.
\end{enumerate}
Following this procedure, it is easy to verify that there is viable parameter space for both $S_1$ and $R_2$ models under our prescribed coupling ansatz, capable of explaining the muon and electron MDM results -- regardless of whether we take the Cs or Rb result to be true. Taking these results together with Figure~\ref{fig:MuonMass} (c), we can infer the predicted muon EDM from the determined viable parameters.

\section{A brief comment on the third lepton flavour}  
\label{sec:ThirdFamily}
One may notice at this point that there has been no mention of the couplings to the tau lepton. This is because the precision with which the tau MDM can be measured is not yet sensitive enough to be directly contrasted with the SM prediction~\cite{taummrev}.

Nevertheless,  consider the case that a LQ coupling exists between the tau lepton and either the charm or the top quark. If it were to couple directly to the up quark, we would expect it would rapidly approach bounds from couplings to the first generation of quarks, however the below bounds would not be applicable.

Considering first the case that the tau lepton couples solely to the top quark. We note that LQ corrections to the branching ratio Br$(\tau \to \mu\gamma)$ are also generated in the case of nonzero $\Delta a_\tau$~\cite{Crivellin:2018qmi} such that
\begin{equation}
    \text{Br}[\tau \to \mu \gamma]= \frac{\alpha }{16 \Gamma_\tau} \frac{m_\tau^2}{m_\mu}|\Delta a_\mu \Delta a_\tau|< 4.2 \times 10^{-8},
\end{equation}
where we have sourced the upper-bound on $\text{Br}[\tau \to \mu \gamma]$ from the Belle collaboration~\cite{Gherardi:2020qhc}. This implies the following bound on $\Delta a_\tau$, if the experimental value of $\Delta a_\mu$ remains at its current value:
\begin{equation}
|\Delta a_\tau|_{\rm{top}}< 2.5 \times 10^{-9}
\end{equation}

Alternatively, if the LQ model couples the tau solely to the charm quark, we can relate the branching ratio Br$(\tau \to e\gamma)$ to $\Delta a_e$ and $\Delta a_\tau$, such that
\begin{equation}
    \text{Br}[\tau \to e \gamma]= \frac{\alpha }{16 \Gamma_\tau} \frac{m_\tau^2}{m_e}|\Delta a_e \Delta a_\tau|< 3.3 \times 10^{-8},
\end{equation}
again sourcing the upper-bound from the Belle collaboration~\cite{Gherardi:2020qhc}. Assuming that $|\Delta a_e|$ remains of order $10^{-13}$, this implies that
\begin{equation}
|\Delta a_\tau|_{\rm{charm}} \lesssim 2.3 \times 10^{-7}.
\end{equation}
The results above show that if an anomaly in the tau MDM is established, then these single LQ solutions could be extended to the third family, and be probed using charged-lepton flavour violating decays. These results can be used to guide model building with scalar LQs in the context of lepton flavour violation. Tau leptons also have strong prospects for the measurement of CP-violating observables on the horizon~\cite{Belle-II:2018jsg}, which could lead to an interesting study of parameter space considering the implications of complex LQ couplings in this sector -- see, for example, references~\cite{Bernreuther:2021elu,Crivellin:2019qnh}.

\section{Conclusions}
\label{sec:conclusion}
In this paper, we have studied the two mixed-chiral scalar leptoquarks $S_1$ and $R_2$, and the interplay between their contributions to MDMs, EDMs and radiative charged-lepton mass corrections. For each LQ, we adopted a flavour ansatz such that the LQ's contribution to the muon dipole operator was generated (at leading-order) by a top-containing loop, and the electron by a charm-containing loop. We allowed LQ couplings to take complex values, and thus built upon the parameter space studied under a similar framework in reference~\cite{Bigaran:2019bqv}. This extension of parameter space was represented via the difference in phase between LH and RH electron-charm~($\xi_{12}$) and muon-top~($\xi_{23}$)  couplings -- a quantity directly probed by the measurement of lepton electric dipole moments. For the electron-charm sector, current experimental constraints prefer a phase difference very close to $\xi_{12}=0$ or $\pi$. For this reason, we restricted electron-charm couplings to real values.

We also addressed a common critique of these chirally-enhanced models for lepton dipole moments, specifically the fine-tuning required to suppress radiative corrections to the associated lepton masses. To quantify what we assessed  not to be fine-tuned, we imposed the requirement that the magnitude of the LQ correction not exceed the lepton pole mass. Using this, we argued that the corrections to the electron mass considered together with satisfying either Cs or Rb MDM results, led to an upper bound for either LQ mass of \begin{align}m_\phi \lesssim 3.12\;\text{TeV}.\end{align} The combination of muon mass constraint and the muon MDM generated weaker bounds on the LQ mass, but these were provided as they are applicable for any $R_2$ or $S_1$ models with top-mass enhanced contributions to $\Delta a_\mu$, such as have become common in the literature. 

Utilising the the fine-tuning restriction on the muon mass, we derived indicative upper-bounds on the magnitude of the muon EDM generated by our models, \begin{align}
|d_\mu|< \mathcal{O}(10^{-22})\; e\;{\rm cm},
\end{align}
which are within reach of future experiments, such as muEDM at PSI, and even FNAL Muon g$-$2 (depending on the value of $\xi_{23}$). The extension of these simple scalar LQ models to address MDM anomalies with CP-violating couplings can thus be probed directly in the near future.

We also performed a comprehensive study of parameter constraints, including those from electroweak processes, the kaon sector, and high $p_T$ dilepton tails. Our results establish that both $S_1$ and $R_2$ remain viable BSM candidates for explaining both $\Delta a_\mu$ and $\Delta a_e $ anomalies, regardless of whether the results for Cs or Rb are confirmed.
The future of muon EDM experiments provide exciting prospects for probing such models in the case of nonzero CP violation in the LQ coupling matrices.

Note that the structure of preferred nonzero LQ couplings in the textures proposed in this work could be motivated by flavour symmetry models. Approaches can be explored to generate such coupling structures in UV-complete BSM models; for example, Froggatt-Nielson mechanisms~\cite{FROGGATT1979277} . Embedding these textures in a UV-complete model is beyond the scope of this work, although we identify it as an avenue for future exploration.

Considering our analysis in the context of wider flavour studies, we motivated future exploration of these LQs' novel parameter spaces. We addressed the viability of extensions to our outlined models for explaining potential anomalies in the $\tau$ MDM, if these were to be measured with precision in future experimental programs. Our work here can be straightforwardly incorporated into broader studies of lepton-flavour violation in scalar LQ models.

\begin{acknowledgments}
IB would like to thank Daniel Flynn and Harry M. Quiney for useful discussion about molecular EDMs,   Peter Athron and Michael A. Schmidt for helpful dialogue about the lepton mass corrections, Martin Hoferichter and 
Olcyr Sumensari for their advice on the study of $Z$ effective couplings. This work is supported in part by the Australian Research Council and the Australian Government Research Training Program Scholarship initiative. 
\end{acknowledgments}

\bibliographystyle{aapmrev4-2}
\bibliography{manuscript}

\providecommand{\noopsort}[1]{}\providecommand{\singleletter}[1]{#1}%
\begin{thebibliography}{89}%
\makeatletter
\providecommand \@ifxundefined [1]{%
 \@ifx{#1\undefined}
}%
\providecommand \@ifnum [1]{%
 \ifnum #1\expandafter \@firstoftwo
 \else \expandafter \@secondoftwo
 \fi
}%
\providecommand \@ifx [1]{%
 \ifx #1\expandafter \@firstoftwo
 \else \expandafter \@secondoftwo
 \fi
}%
\providecommand \natexlab [1]{#1}%
\providecommand \enquote  [1]{``#1''}%
\providecommand \bibnamefont  [1]{#1}%
\providecommand \bibfnamefont [1]{#1}%
\providecommand \citenamefont [1]{#1}%
\providecommand \href@noop [0]{\@secondoftwo}%
\providecommand \href [0]{\begingroup \@sanitize@url \@href}%
\providecommand \@href[1]{\@@startlink{#1}\@@href}%
\providecommand \@@href[1]{\endgroup#1\@@endlink}%
\providecommand \@sanitize@url [0]{\catcode `\\12\catcode `\$12\catcode
  `\&12\catcode `\#12\catcode `\^12\catcode `\_12\catcode `\%12\relax}%
\providecommand \@@startlink[1]{}%
\providecommand \@@endlink[0]{}%
\providecommand \url  [0]{\begingroup\@sanitize@url \@url }%
\providecommand \@url [1]{\endgroup\@href {#1}{\urlprefix }}%
\providecommand \urlprefix  [0]{URL }%
\providecommand \Eprint [0]{\href }%
\providecommand \doibase [0]{https://doi.org/}%
\providecommand \selectlanguage [0]{\@gobble}%
\providecommand \bibinfo  [0]{\@secondoftwo}%
\providecommand \bibfield  [0]{\@secondoftwo}%
\providecommand \translation [1]{[#1]}%
\providecommand \BibitemOpen [0]{}%
\providecommand \bibitemStop [0]{}%
\providecommand \bibitemNoStop [0]{.\EOS\space}%
\providecommand \EOS [0]{\spacefactor3000\relax}%
\providecommand \BibitemShut  [1]{\csname bibitem#1\endcsname}%
\let\auto@bib@innerbib\@empty
\bibitem [{\citenamefont {Abi}\ \emph {et~al.}(2021)\citenamefont {Abi} \emph
  {et~al.}}]{Abi:2021gix}%
  \BibitemOpen
  \bibfield  {author} {\bibinfo {author} {\bibfnamefont {B.}~\bibnamefont
  {Abi}} \emph {et~al.} (\bibinfo {collaboration} {Muon g-2}),\ }\href
  {https://doi.org/10.1103/PhysRevLett.126.141801} {\bibfield  {journal}
  {\bibinfo  {journal} {Phys. Rev. Lett.}\ }\textbf {\bibinfo {volume} {126}},\
  \bibinfo {pages} {141801} (\bibinfo {year} {2021})},\ \Eprint
  {https://arxiv.org/abs/2104.03281} {arXiv:2104.03281 [hep-ex]} \BibitemShut
  {NoStop}%
\bibitem [{\citenamefont {Bennett}\ \emph {et~al.}(2004)\citenamefont {Bennett}
  \emph {et~al.}}]{Muong-2:2004fok}%
  \BibitemOpen
  \bibfield  {author} {\bibinfo {author} {\bibfnamefont {G.~W.}\ \bibnamefont
  {Bennett}} \emph {et~al.} (\bibinfo {collaboration} {Muon g-2}),\ }\href
  {https://doi.org/10.1103/PhysRevLett.92.161802} {\bibfield  {journal}
  {\bibinfo  {journal} {Phys. Rev. Lett.}\ }\textbf {\bibinfo {volume} {92}},\
  \bibinfo {pages} {161802} (\bibinfo {year} {2004})},\ \Eprint
  {https://arxiv.org/abs/hep-ex/0401008} {arXiv:hep-ex/0401008} \BibitemShut
  {NoStop}%
\bibitem [{\citenamefont {Aoyama}\ \emph {et~al.}(2020)\citenamefont {Aoyama}
  \emph {et~al.}}]{Aoyama:2020ynm}%
  \BibitemOpen
  \bibfield  {author} {\bibinfo {author} {\bibfnamefont {T.}~\bibnamefont
  {Aoyama}} \emph {et~al.},\ }\href
  {https://doi.org/10.1016/j.physrep.2020.07.006} {\bibfield  {journal}
  {\bibinfo  {journal} {Phys. Rept.}\ }\textbf {\bibinfo {volume} {887}},\
  \bibinfo {pages} {1} (\bibinfo {year} {2020})},\ \Eprint
  {https://arxiv.org/abs/2006.04822} {arXiv:2006.04822 [hep-ph]} \BibitemShut
  {NoStop}%
\bibitem [{\citenamefont {Aoyama}\ \emph {et~al.}(2012)\citenamefont {Aoyama},
  \citenamefont {Hayakawa}, \citenamefont {Kinoshita},\ and\ \citenamefont
  {Nio}}]{aoyama:2012wk}%
  \BibitemOpen
  \bibfield  {author} {\bibinfo {author} {\bibfnamefont {T.}~\bibnamefont
  {Aoyama}}, \bibinfo {author} {\bibfnamefont {M.}~\bibnamefont {Hayakawa}},
  \bibinfo {author} {\bibfnamefont {T.}~\bibnamefont {Kinoshita}},\ and\
  \bibinfo {author} {\bibfnamefont {M.}~\bibnamefont {Nio}},\ }\href
  {https://doi.org/10.1103/PhysRevLett.109.111808} {\bibfield  {journal}
  {\bibinfo  {journal} {Phys. Rev. Lett.}\ }\textbf {\bibinfo {volume} {109}},\
  \bibinfo {pages} {111808} (\bibinfo {year} {2012})},\ \Eprint
  {https://arxiv.org/abs/1205.5370} {arXiv:1205.5370 [hep-ph]} \BibitemShut
  {NoStop}%
\bibitem [{\citenamefont {Aoyama}, \citenamefont {Kinoshita},\ and\
  \citenamefont {Nio}(2019)}]{Aoyama:2019ryr}%
  \BibitemOpen
  \bibfield  {author} {\bibinfo {author} {\bibfnamefont {T.}~\bibnamefont
  {Aoyama}}, \bibinfo {author} {\bibfnamefont {T.}~\bibnamefont {Kinoshita}},\
  and\ \bibinfo {author} {\bibfnamefont {M.}~\bibnamefont {Nio}},\ }\href
  {https://doi.org/10.3390/atoms7010028} {\bibfield  {journal} {\bibinfo
  {journal} {Atoms}\ }\textbf {\bibinfo {volume} {7}},\ \bibinfo {pages} {28}
  (\bibinfo {year} {2019})}\BibitemShut {NoStop}%
\bibitem [{\citenamefont {Czarnecki}, \citenamefont {Marciano},\ and\
  \citenamefont {Vainshtein}(2003)}]{czarnecki:2002nt}%
  \BibitemOpen
  \bibfield  {author} {\bibinfo {author} {\bibfnamefont {A.}~\bibnamefont
  {Czarnecki}}, \bibinfo {author} {\bibfnamefont {W.~J.}\ \bibnamefont
  {Marciano}},\ and\ \bibinfo {author} {\bibfnamefont {A.}~\bibnamefont
  {Vainshtein}},\ }\href {https://doi.org/10.1103/PhysRevD.67.073006}
  {\bibfield  {journal} {\bibinfo  {journal} {Phys. Rev.}\ }\textbf {\bibinfo
  {volume} {D67}},\ \bibinfo {pages} {073006} (\bibinfo {year} {2003})},\
  \bibinfo {note} {[Erratum: Phys. Rev. {\bf D73}, 119901 (2006)]},\ \Eprint
  {https://arxiv.org/abs/hep-ph/0212229} {arXiv:hep-ph/0212229 [hep-ph]}
  \BibitemShut {NoStop}%
\bibitem [{\citenamefont {Gnendiger}, \citenamefont {St{\"o}ckinger},\ and\
  \citenamefont {St{\"o}ckinger-Kim}(2013)}]{gnendiger:2013pva}%
  \BibitemOpen
  \bibfield  {author} {\bibinfo {author} {\bibfnamefont {C.}~\bibnamefont
  {Gnendiger}}, \bibinfo {author} {\bibfnamefont {D.}~\bibnamefont
  {St{\"o}ckinger}},\ and\ \bibinfo {author} {\bibfnamefont {H.}~\bibnamefont
  {St{\"o}ckinger-Kim}},\ }\href {https://doi.org/10.1103/PhysRevD.88.053005}
  {\bibfield  {journal} {\bibinfo  {journal} {Phys. Rev.}\ }\textbf {\bibinfo
  {volume} {D88}},\ \bibinfo {pages} {053005} (\bibinfo {year} {2013})},\
  \Eprint {https://arxiv.org/abs/1306.5546} {arXiv:1306.5546 [hep-ph]}
  \BibitemShut {NoStop}%
\bibitem [{\citenamefont {Davier}\ \emph {et~al.}(2017)\citenamefont {Davier},
  \citenamefont {Hoecker}, \citenamefont {Malaescu},\ and\ \citenamefont
  {Zhang}}]{davier:2017zfy}%
  \BibitemOpen
  \bibfield  {author} {\bibinfo {author} {\bibfnamefont {M.}~\bibnamefont
  {Davier}}, \bibinfo {author} {\bibfnamefont {A.}~\bibnamefont {Hoecker}},
  \bibinfo {author} {\bibfnamefont {B.}~\bibnamefont {Malaescu}},\ and\
  \bibinfo {author} {\bibfnamefont {Z.}~\bibnamefont {Zhang}},\ }\href
  {https://doi.org/10.1140/epjc/s10052-017-5161-6} {\bibfield  {journal}
  {\bibinfo  {journal} {Eur. Phys. J.}\ }\textbf {\bibinfo {volume} {C77}},\
  \bibinfo {pages} {827} (\bibinfo {year} {2017})},\ \Eprint
  {https://arxiv.org/abs/1706.09436} {arXiv:1706.09436 [hep-ph]} \BibitemShut
  {NoStop}%
\bibitem [{\citenamefont {Keshavarzi}, \citenamefont {Nomura},\ and\
  \citenamefont {Teubner}(2018)}]{keshavarzi:2018mgv}%
  \BibitemOpen
  \bibfield  {author} {\bibinfo {author} {\bibfnamefont {A.}~\bibnamefont
  {Keshavarzi}}, \bibinfo {author} {\bibfnamefont {D.}~\bibnamefont {Nomura}},\
  and\ \bibinfo {author} {\bibfnamefont {T.}~\bibnamefont {Teubner}},\ }\href
  {https://doi.org/10.1103/PhysRevD.97.114025} {\bibfield  {journal} {\bibinfo
  {journal} {Phys. Rev.}\ }\textbf {\bibinfo {volume} {D97}},\ \bibinfo {pages}
  {114025} (\bibinfo {year} {2018})},\ \Eprint
  {https://arxiv.org/abs/1802.02995} {arXiv:1802.02995 [hep-ph]} \BibitemShut
  {NoStop}%
\bibitem [{\citenamefont {Colangelo}, \citenamefont {Hoferichter},\ and\
  \citenamefont {Stoffer}(2019)}]{colangelo:2018mtw}%
  \BibitemOpen
  \bibfield  {author} {\bibinfo {author} {\bibfnamefont {G.}~\bibnamefont
  {Colangelo}}, \bibinfo {author} {\bibfnamefont {M.}~\bibnamefont
  {Hoferichter}},\ and\ \bibinfo {author} {\bibfnamefont {P.}~\bibnamefont
  {Stoffer}},\ }\href {https://doi.org/10.1007/JHEP02(2019)006} {\bibfield
  {journal} {\bibinfo  {journal} {JHEP}\ }\textbf {\bibinfo {volume} {02}},\
  \bibinfo {pages} {006} (\bibinfo {year} {2019})},\ \Eprint
  {https://arxiv.org/abs/1810.00007} {arXiv:1810.00007 [hep-ph]} \BibitemShut
  {NoStop}%
\bibitem [{\citenamefont {Hoferichter}, \citenamefont {Hoid},\ and\
  \citenamefont {Kubis}(2019)}]{hoferichter:2019gzf}%
  \BibitemOpen
  \bibfield  {author} {\bibinfo {author} {\bibfnamefont {M.}~\bibnamefont
  {Hoferichter}}, \bibinfo {author} {\bibfnamefont {B.-L.}\ \bibnamefont
  {Hoid}},\ and\ \bibinfo {author} {\bibfnamefont {B.}~\bibnamefont {Kubis}},\
  }\href {https://doi.org/10.1007/JHEP08(2019)137} {\bibfield  {journal}
  {\bibinfo  {journal} {JHEP}\ }\textbf {\bibinfo {volume} {08}},\ \bibinfo
  {pages} {137} (\bibinfo {year} {2019})},\ \Eprint
  {https://arxiv.org/abs/1907.01556} {arXiv:1907.01556 [hep-ph]} \BibitemShut
  {NoStop}%
\bibitem [{\citenamefont {Davier}\ \emph {et~al.}(2020)\citenamefont {Davier},
  \citenamefont {Hoecker}, \citenamefont {Malaescu},\ and\ \citenamefont
  {Zhang}}]{davier:2019can}%
  \BibitemOpen
  \bibfield  {author} {\bibinfo {author} {\bibfnamefont {M.}~\bibnamefont
  {Davier}}, \bibinfo {author} {\bibfnamefont {A.}~\bibnamefont {Hoecker}},
  \bibinfo {author} {\bibfnamefont {B.}~\bibnamefont {Malaescu}},\ and\
  \bibinfo {author} {\bibfnamefont {Z.}~\bibnamefont {Zhang}},\ }\href
  {https://doi.org/10.1140/epjc/s10052-020-7792-2} {\bibfield  {journal}
  {\bibinfo  {journal} {Eur. Phys. J.}\ }\textbf {\bibinfo {volume} {C80}},\
  \bibinfo {pages} {241} (\bibinfo {year} {2020})},\ \bibinfo {note} {[Erratum:
  Eur. Phys. J. {\bf C80}, 410 (2020)]},\ \Eprint
  {https://arxiv.org/abs/1908.00921} {arXiv:1908.00921 [hep-ph]} \BibitemShut
  {NoStop}%
\bibitem [{\citenamefont {Keshavarzi}, \citenamefont {Nomura},\ and\
  \citenamefont {Teubner}(2020)}]{keshavarzi:2019abf}%
  \BibitemOpen
  \bibfield  {author} {\bibinfo {author} {\bibfnamefont {A.}~\bibnamefont
  {Keshavarzi}}, \bibinfo {author} {\bibfnamefont {D.}~\bibnamefont {Nomura}},\
  and\ \bibinfo {author} {\bibfnamefont {T.}~\bibnamefont {Teubner}},\ }\href
  {https://doi.org/10.1103/PhysRevD.101.014029} {\bibfield  {journal} {\bibinfo
   {journal} {Phys. Rev.}\ }\textbf {\bibinfo {volume} {D101}},\ \bibinfo
  {pages} {014029} (\bibinfo {year} {2020})},\ \Eprint
  {https://arxiv.org/abs/1911.00367} {arXiv:1911.00367 [hep-ph]} \BibitemShut
  {NoStop}%
\bibitem [{\citenamefont {Kurz}\ \emph {et~al.}(2014)\citenamefont {Kurz},
  \citenamefont {Liu}, \citenamefont {Marquard},\ and\ \citenamefont
  {Steinhauser}}]{kurz:2014wya}%
  \BibitemOpen
  \bibfield  {author} {\bibinfo {author} {\bibfnamefont {A.}~\bibnamefont
  {Kurz}}, \bibinfo {author} {\bibfnamefont {T.}~\bibnamefont {Liu}}, \bibinfo
  {author} {\bibfnamefont {P.}~\bibnamefont {Marquard}},\ and\ \bibinfo
  {author} {\bibfnamefont {M.}~\bibnamefont {Steinhauser}},\ }\href
  {https://doi.org/10.1016/j.physletb.2014.05.043} {\bibfield  {journal}
  {\bibinfo  {journal} {Phys. Lett.}\ }\textbf {\bibinfo {volume} {B734}},\
  \bibinfo {pages} {144} (\bibinfo {year} {2014})},\ \Eprint
  {https://arxiv.org/abs/1403.6400} {arXiv:1403.6400 [hep-ph]} \BibitemShut
  {NoStop}%
\bibitem [{\citenamefont {Melnikov}\ and\ \citenamefont
  {Vainshtein}(2004)}]{melnikov:2003xd}%
  \BibitemOpen
  \bibfield  {author} {\bibinfo {author} {\bibfnamefont {K.}~\bibnamefont
  {Melnikov}}\ and\ \bibinfo {author} {\bibfnamefont {A.}~\bibnamefont
  {Vainshtein}},\ }\href {https://doi.org/10.1103/PhysRevD.70.113006}
  {\bibfield  {journal} {\bibinfo  {journal} {Phys. Rev.}\ }\textbf {\bibinfo
  {volume} {D70}},\ \bibinfo {pages} {113006} (\bibinfo {year} {2004})},\
  \Eprint {https://arxiv.org/abs/hep-ph/0312226} {arXiv:hep-ph/0312226
  [hep-ph]} \BibitemShut {NoStop}%
\bibitem [{\citenamefont {Masjuan}\ and\ \citenamefont
  {S{\'a}nchez-Puertas}(2017)}]{masjuan:2017tvw}%
  \BibitemOpen
  \bibfield  {author} {\bibinfo {author} {\bibfnamefont {P.}~\bibnamefont
  {Masjuan}}\ and\ \bibinfo {author} {\bibfnamefont {P.}~\bibnamefont
  {S{\'a}nchez-Puertas}},\ }\href {https://doi.org/10.1103/PhysRevD.95.054026}
  {\bibfield  {journal} {\bibinfo  {journal} {Phys. Rev.}\ }\textbf {\bibinfo
  {volume} {D95}},\ \bibinfo {pages} {054026} (\bibinfo {year} {2017})},\
  \Eprint {https://arxiv.org/abs/1701.05829} {arXiv:1701.05829 [hep-ph]}
  \BibitemShut {NoStop}%
\bibitem [{\citenamefont {Colangelo}\ \emph {et~al.}(2017)\citenamefont
  {Colangelo}, \citenamefont {Hoferichter}, \citenamefont {Procura},\ and\
  \citenamefont {Stoffer}}]{Colangelo:2017fiz}%
  \BibitemOpen
  \bibfield  {author} {\bibinfo {author} {\bibfnamefont {G.}~\bibnamefont
  {Colangelo}}, \bibinfo {author} {\bibfnamefont {M.}~\bibnamefont
  {Hoferichter}}, \bibinfo {author} {\bibfnamefont {M.}~\bibnamefont
  {Procura}},\ and\ \bibinfo {author} {\bibfnamefont {P.}~\bibnamefont
  {Stoffer}},\ }\href {https://doi.org/10.1007/JHEP04(2017)161} {\bibfield
  {journal} {\bibinfo  {journal} {JHEP}\ }\textbf {\bibinfo {volume} {04}},\
  \bibinfo {pages} {161} (\bibinfo {year} {2017})},\ \Eprint
  {https://arxiv.org/abs/1702.07347} {arXiv:1702.07347 [hep-ph]} \BibitemShut
  {NoStop}%
\bibitem [{\citenamefont {Hoferichter}\ \emph {et~al.}(2018)\citenamefont
  {Hoferichter}, \citenamefont {Hoid}, \citenamefont {Kubis}, \citenamefont
  {Leupold},\ and\ \citenamefont {Schneider}}]{hoferichter:2018kwz}%
  \BibitemOpen
  \bibfield  {author} {\bibinfo {author} {\bibfnamefont {M.}~\bibnamefont
  {Hoferichter}}, \bibinfo {author} {\bibfnamefont {B.-L.}\ \bibnamefont
  {Hoid}}, \bibinfo {author} {\bibfnamefont {B.}~\bibnamefont {Kubis}},
  \bibinfo {author} {\bibfnamefont {S.}~\bibnamefont {Leupold}},\ and\ \bibinfo
  {author} {\bibfnamefont {S.~P.}\ \bibnamefont {Schneider}},\ }\href
  {https://doi.org/10.1007/JHEP10(2018)141} {\bibfield  {journal} {\bibinfo
  {journal} {JHEP}\ }\textbf {\bibinfo {volume} {10}},\ \bibinfo {pages} {141}
  (\bibinfo {year} {2018})},\ \Eprint {https://arxiv.org/abs/1808.04823}
  {arXiv:1808.04823 [hep-ph]} \BibitemShut {NoStop}%
\bibitem [{\citenamefont {G{\'e}rardin}, \citenamefont {Meyer},\ and\
  \citenamefont {Nyffeler}(2019)}]{gerardin:2019vio}%
  \BibitemOpen
  \bibfield  {author} {\bibinfo {author} {\bibfnamefont {A.}~\bibnamefont
  {G{\'e}rardin}}, \bibinfo {author} {\bibfnamefont {H.~B.}\ \bibnamefont
  {Meyer}},\ and\ \bibinfo {author} {\bibfnamefont {A.}~\bibnamefont
  {Nyffeler}},\ }\href {https://doi.org/10.1103/PhysRevD.100.034520} {\bibfield
   {journal} {\bibinfo  {journal} {Phys. Rev.}\ }\textbf {\bibinfo {volume}
  {D100}},\ \bibinfo {pages} {034520} (\bibinfo {year} {2019})},\ \Eprint
  {https://arxiv.org/abs/1903.09471} {arXiv:1903.09471 [hep-lat]} \BibitemShut
  {NoStop}%
\bibitem [{\citenamefont {Bijnens}, \citenamefont {Hermansson-Truedsson},\ and\
  \citenamefont {Rodr{\'i}guez-S{\'a}nchez}(2019)}]{bijnens:2019ghy}%
  \BibitemOpen
  \bibfield  {author} {\bibinfo {author} {\bibfnamefont {J.}~\bibnamefont
  {Bijnens}}, \bibinfo {author} {\bibfnamefont {N.}~\bibnamefont
  {Hermansson-Truedsson}},\ and\ \bibinfo {author} {\bibfnamefont
  {A.}~\bibnamefont {Rodr{\'i}guez-S{\'a}nchez}},\ }\href
  {https://doi.org/10.1016/j.physletb.2019.134994} {\bibfield  {journal}
  {\bibinfo  {journal} {Phys. Lett.}\ }\textbf {\bibinfo {volume} {B798}},\
  \bibinfo {pages} {134994} (\bibinfo {year} {2019})},\ \Eprint
  {https://arxiv.org/abs/1908.03331} {arXiv:1908.03331 [hep-ph]} \BibitemShut
  {NoStop}%
\bibitem [{\citenamefont {Colangelo}\ \emph {et~al.}(2020)\citenamefont
  {Colangelo}, \citenamefont {Hagelstein}, \citenamefont {Hoferichter},
  \citenamefont {Laub},\ and\ \citenamefont {Stoffer}}]{colangelo:2019uex}%
  \BibitemOpen
  \bibfield  {author} {\bibinfo {author} {\bibfnamefont {G.}~\bibnamefont
  {Colangelo}}, \bibinfo {author} {\bibfnamefont {F.}~\bibnamefont
  {Hagelstein}}, \bibinfo {author} {\bibfnamefont {M.}~\bibnamefont
  {Hoferichter}}, \bibinfo {author} {\bibfnamefont {L.}~\bibnamefont {Laub}},\
  and\ \bibinfo {author} {\bibfnamefont {P.}~\bibnamefont {Stoffer}},\ }\href
  {https://doi.org/10.1007/JHEP03(2020)101} {\bibfield  {journal} {\bibinfo
  {journal} {JHEP}\ }\textbf {\bibinfo {volume} {03}},\ \bibinfo {pages} {101}
  (\bibinfo {year} {2020})},\ \Eprint {https://arxiv.org/abs/1910.13432}
  {arXiv:1910.13432 [hep-ph]} \BibitemShut {NoStop}%
\bibitem [{\citenamefont {Blum}\ \emph {et~al.}(2020)\citenamefont {Blum},
  \citenamefont {Christ}, \citenamefont {Hayakawa}, \citenamefont {Izubuchi},
  \citenamefont {Jin}, \citenamefont {Jung},\ and\ \citenamefont
  {Lehner}}]{Blum:2019ugy}%
  \BibitemOpen
  \bibfield  {author} {\bibinfo {author} {\bibfnamefont {T.}~\bibnamefont
  {Blum}}, \bibinfo {author} {\bibfnamefont {N.}~\bibnamefont {Christ}},
  \bibinfo {author} {\bibfnamefont {M.}~\bibnamefont {Hayakawa}}, \bibinfo
  {author} {\bibfnamefont {T.}~\bibnamefont {Izubuchi}}, \bibinfo {author}
  {\bibfnamefont {L.}~\bibnamefont {Jin}}, \bibinfo {author} {\bibfnamefont
  {C.}~\bibnamefont {Jung}},\ and\ \bibinfo {author} {\bibfnamefont
  {C.}~\bibnamefont {Lehner}},\ }\href
  {https://doi.org/10.1103/PhysRevLett.124.132002} {\bibfield  {journal}
  {\bibinfo  {journal} {Phys. Rev. Lett.}\ }\textbf {\bibinfo {volume} {124}},\
  \bibinfo {pages} {132002} (\bibinfo {year} {2020})},\ \Eprint
  {https://arxiv.org/abs/1911.08123} {arXiv:1911.08123 [hep-lat]} \BibitemShut
  {NoStop}%
\bibitem [{\citenamefont {Colangelo}\ \emph {et~al.}(2014)\citenamefont
  {Colangelo}, \citenamefont {Hoferichter}, \citenamefont {Nyffeler},
  \citenamefont {Passera},\ and\ \citenamefont {Stoffer}}]{colangelo:2014qya}%
  \BibitemOpen
  \bibfield  {author} {\bibinfo {author} {\bibfnamefont {G.}~\bibnamefont
  {Colangelo}}, \bibinfo {author} {\bibfnamefont {M.}~\bibnamefont
  {Hoferichter}}, \bibinfo {author} {\bibfnamefont {A.}~\bibnamefont
  {Nyffeler}}, \bibinfo {author} {\bibfnamefont {M.}~\bibnamefont {Passera}},\
  and\ \bibinfo {author} {\bibfnamefont {P.}~\bibnamefont {Stoffer}},\ }\href
  {https://doi.org/10.1016/j.physletb.2014.06.012} {\bibfield  {journal}
  {\bibinfo  {journal} {Phys. Lett.}\ }\textbf {\bibinfo {volume} {B735}},\
  \bibinfo {pages} {90} (\bibinfo {year} {2014})},\ \Eprint
  {https://arxiv.org/abs/1403.7512} {arXiv:1403.7512 [hep-ph]} \BibitemShut
  {NoStop}%
\bibitem [{\citenamefont {Parker}\ \emph {et~al.}(2018)\citenamefont {Parker},
  \citenamefont {Yu}, \citenamefont {Zhong}, \citenamefont {Estey},\ and\
  \citenamefont {M\~aeller}}]{articleParker}%
  \BibitemOpen
  \bibfield  {author} {\bibinfo {author} {\bibfnamefont {R.}~\bibnamefont
  {Parker}}, \bibinfo {author} {\bibfnamefont {C.}~\bibnamefont {Yu}}, \bibinfo
  {author} {\bibfnamefont {W.}~\bibnamefont {Zhong}}, \bibinfo {author}
  {\bibfnamefont {B.}~\bibnamefont {Estey}},\ and\ \bibinfo {author}
  {\bibfnamefont {H.}~\bibnamefont {M\~aeller}},\ }\href
  {https://doi.org/10.1126/science.aap7706} {\bibfield  {journal} {\bibinfo
  {journal} {Science}\ }\textbf {\bibinfo {volume} {360}},\ \bibinfo {pages}
  {191} (\bibinfo {year} {2018})}\BibitemShut {NoStop}%
\bibitem [{\citenamefont {Morel}\ \emph {et~al.}(2020)\citenamefont {Morel},
  \citenamefont {Yao}, \citenamefont {Clad\'e},\ and\ \citenamefont
  {Guellati-Kh\'elifa}}]{Morel:2020dww}%
  \BibitemOpen
  \bibfield  {author} {\bibinfo {author} {\bibfnamefont {L.}~\bibnamefont
  {Morel}}, \bibinfo {author} {\bibfnamefont {Z.}~\bibnamefont {Yao}}, \bibinfo
  {author} {\bibfnamefont {P.}~\bibnamefont {Clad\'e}},\ and\ \bibinfo {author}
  {\bibfnamefont {S.}~\bibnamefont {Guellati-Kh\'elifa}},\ }\href
  {https://doi.org/10.1038/s41586-020-2964-7} {\bibfield  {journal} {\bibinfo
  {journal} {Nature}\ }\textbf {\bibinfo {volume} {588}},\ \bibinfo {pages}
  {61} (\bibinfo {year} {2020})}\BibitemShut {NoStop}%
\bibitem [{\citenamefont {Bennett}\ \emph {et~al.}(2009)\citenamefont {Bennett}
  \emph {et~al.}}]{Bennett:2008dy}%
  \BibitemOpen
  \bibfield  {author} {\bibinfo {author} {\bibfnamefont {G.}~\bibnamefont
  {Bennett}} \emph {et~al.} (\bibinfo {collaboration} {Muon (g-2)}),\ }\href
  {https://doi.org/10.1103/PhysRevD.80.052008} {\bibfield  {journal} {\bibinfo
  {journal} {Phys. Rev. D}\ }\textbf {\bibinfo {volume} {80}},\ \bibinfo
  {pages} {052008} (\bibinfo {year} {2009})},\ \Eprint
  {https://arxiv.org/abs/0811.1207} {arXiv:0811.1207 [hep-ex]} \BibitemShut
  {NoStop}%
\bibitem [{\citenamefont {Andreev}\ \emph {et~al.}(2018)\citenamefont {Andreev}
  \emph {et~al.}}]{Andreev:2018ayy}%
  \BibitemOpen
  \bibfield  {author} {\bibinfo {author} {\bibfnamefont {V.}~\bibnamefont
  {Andreev}} \emph {et~al.} (\bibinfo {collaboration} {ACME}),\ }\href
  {https://doi.org/10.1038/s41586-018-0599-8} {\bibfield  {journal} {\bibinfo
  {journal} {Nature}\ }\textbf {\bibinfo {volume} {562}},\ \bibinfo {pages}
  {355} (\bibinfo {year} {2018})}\BibitemShut {NoStop}%
\bibitem [{\citenamefont {Bigaran}\ and\ \citenamefont
  {Volkas}(2020)}]{Bigaran:2020jil}%
  \BibitemOpen
  \bibfield  {author} {\bibinfo {author} {\bibfnamefont {I.}~\bibnamefont
  {Bigaran}}\ and\ \bibinfo {author} {\bibfnamefont {R.~R.}\ \bibnamefont
  {Volkas}},\ }\href {https://doi.org/10.1103/PhysRevD.102.075037} {\bibfield
  {journal} {\bibinfo  {journal} {Phys. Rev. D}\ }\textbf {\bibinfo {volume}
  {102}},\ \bibinfo {pages} {075037} (\bibinfo {year} {2020})},\ \Eprint
  {https://arxiv.org/abs/2002.12544} {arXiv:2002.12544 [hep-ph]} \BibitemShut
  {NoStop}%
\bibitem [{\citenamefont {Crivellin}, \citenamefont {Hoferichter},\ and\
  \citenamefont {Schmidt-Wellenburg}(2018)}]{Crivellin:2018qmi}%
  \BibitemOpen
  \bibfield  {author} {\bibinfo {author} {\bibfnamefont {A.}~\bibnamefont
  {Crivellin}}, \bibinfo {author} {\bibfnamefont {M.}~\bibnamefont
  {Hoferichter}},\ and\ \bibinfo {author} {\bibfnamefont {P.}~\bibnamefont
  {Schmidt-Wellenburg}},\ }\href {https://doi.org/10.1103/PhysRevD.98.113002}
  {\bibfield  {journal} {\bibinfo  {journal} {Phys. Rev.}\ }\textbf {\bibinfo
  {volume} {D98}},\ \bibinfo {pages} {113002} (\bibinfo {year} {2018})},\
  \Eprint {https://arxiv.org/abs/1807.11484} {arXiv:1807.11484 [hep-ph]}
  \BibitemShut {NoStop}%
\bibitem [{\citenamefont {Dekens}\ \emph {et~al.}(2019)\citenamefont {Dekens},
  \citenamefont {de~Vries}, \citenamefont {Jung},\ and\ \citenamefont
  {Vos}}]{Dekens:2018bci}%
  \BibitemOpen
  \bibfield  {author} {\bibinfo {author} {\bibfnamefont {W.}~\bibnamefont
  {Dekens}}, \bibinfo {author} {\bibfnamefont {J.}~\bibnamefont {de~Vries}},
  \bibinfo {author} {\bibfnamefont {M.}~\bibnamefont {Jung}},\ and\ \bibinfo
  {author} {\bibfnamefont {K.~K.}\ \bibnamefont {Vos}},\ }\href
  {https://doi.org/10.1007/JHEP01(2019)069} {\bibfield  {journal} {\bibinfo
  {journal} {JHEP}\ }\textbf {\bibinfo {volume} {01}},\ \bibinfo {pages} {069}
  (\bibinfo {year} {2019})},\ \Eprint {https://arxiv.org/abs/1809.09114}
  {arXiv:1809.09114 [hep-ph]} \BibitemShut {NoStop}%
\bibitem [{\citenamefont {Fuyuto}, \citenamefont {Ramsey-Musolf},\ and\
  \citenamefont {Shen}(2019)}]{Fuyuto:2018scm}%
  \BibitemOpen
  \bibfield  {author} {\bibinfo {author} {\bibfnamefont {K.}~\bibnamefont
  {Fuyuto}}, \bibinfo {author} {\bibfnamefont {M.}~\bibnamefont
  {Ramsey-Musolf}},\ and\ \bibinfo {author} {\bibfnamefont {T.}~\bibnamefont
  {Shen}},\ }\href {https://doi.org/10.1016/j.physletb.2018.11.016} {\bibfield
  {journal} {\bibinfo  {journal} {Phys. Lett. B}\ }\textbf {\bibinfo {volume}
  {788}},\ \bibinfo {pages} {52} (\bibinfo {year} {2019})},\ \Eprint
  {https://arxiv.org/abs/1804.01137} {arXiv:1804.01137 [hep-ph]} \BibitemShut
  {NoStop}%
\bibitem [{\citenamefont {Bernreuther}, \citenamefont {Chen},\ and\
  \citenamefont {Nachtmann}(2021)}]{Bernreuther:2021elu}%
  \BibitemOpen
  \bibfield  {author} {\bibinfo {author} {\bibfnamefont {W.}~\bibnamefont
  {Bernreuther}}, \bibinfo {author} {\bibfnamefont {L.}~\bibnamefont {Chen}},\
  and\ \bibinfo {author} {\bibfnamefont {O.}~\bibnamefont {Nachtmann}},\ }\href
  {https://doi.org/10.1103/PhysRevD.103.096011} {\bibfield  {journal} {\bibinfo
   {journal} {Phys. Rev. D}\ }\textbf {\bibinfo {volume} {103}},\ \bibinfo
  {pages} {096011} (\bibinfo {year} {2021})},\ \Eprint
  {https://arxiv.org/abs/2101.08071} {arXiv:2101.08071 [hep-ph]} \BibitemShut
  {NoStop}%
\bibitem [{\citenamefont {Lee}(2021)}]{Lee:2021jdr}%
  \BibitemOpen
  \bibfield  {author} {\bibinfo {author} {\bibfnamefont {H.~M.}\ \bibnamefont
  {Lee}},\ }\href {https://doi.org/10.1103/PhysRevD.104.015007} {\bibfield
  {journal} {\bibinfo  {journal} {Phys. Rev. D}\ }\textbf {\bibinfo {volume}
  {104}},\ \bibinfo {pages} {015007} (\bibinfo {year} {2021})},\ \Eprint
  {https://arxiv.org/abs/2104.02982} {arXiv:2104.02982 [hep-ph]} \BibitemShut
  {NoStop}%
\bibitem [{\citenamefont {Pati}, \citenamefont {Salam},\ and\ \citenamefont
  {Strathdee}(1975)}]{articlePSFN}%
  \BibitemOpen
  \bibfield  {author} {\bibinfo {author} {\bibfnamefont {J.}~\bibnamefont
  {Pati}}, \bibinfo {author} {\bibfnamefont {A.}~\bibnamefont {Salam}},\ and\
  \bibinfo {author} {\bibfnamefont {J.}~\bibnamefont {Strathdee}},\ }\href
  {https://doi.org/10.1007/BF02849600} {\bibfield  {journal} {\bibinfo
  {journal} {Il Nuovo Cimento A}\ }\textbf {\bibinfo {volume} {26}},\ \bibinfo
  {pages} {72} (\bibinfo {year} {1975})}\BibitemShut {NoStop}%
\bibitem [{\citenamefont {Dor\u{s}ner}\ \emph {et~al.}(2016)\citenamefont
  {Dor\u{s}ner}, \citenamefont {Fajfer}, \citenamefont {Greljo}, \citenamefont
  {Kamenik},\ and\ \citenamefont {Ko\u{s}nik}}]{Dorsner:2016wpm}%
  \BibitemOpen
  \bibfield  {author} {\bibinfo {author} {\bibfnamefont {I.}~\bibnamefont
  {Dor\u{s}ner}}, \bibinfo {author} {\bibfnamefont {S.}~\bibnamefont {Fajfer}},
  \bibinfo {author} {\bibfnamefont {A.}~\bibnamefont {Greljo}}, \bibinfo
  {author} {\bibfnamefont {J.~F.}\ \bibnamefont {Kamenik}},\ and\ \bibinfo
  {author} {\bibfnamefont {N.}~\bibnamefont {Ko\u{s}nik}},\ }\href
  {https://doi.org/10.1016/j.physrep.2016.06.001} {\bibfield  {journal}
  {\bibinfo  {journal} {Phys. Rept.}\ }\textbf {\bibinfo {volume} {641}},\
  \bibinfo {pages} {1} (\bibinfo {year} {2016})},\ \Eprint
  {https://arxiv.org/abs/1603.04993} {arXiv:1603.04993 [hep-ph]} \BibitemShut
  {NoStop}%
\bibitem [{\citenamefont {Be\u{c}irevi\'{c}}\ \emph {et~al.}(2016)\citenamefont
  {Be\u{c}irevi\'{c}}, \citenamefont {Ko\u{s}nik}, \citenamefont {Sumensari},\
  and\ \citenamefont {Zukanovich~Funchal}}]{Becirevic:2016oho}%
  \BibitemOpen
  \bibfield  {author} {\bibinfo {author} {\bibfnamefont {D.}~\bibnamefont
  {Be\u{c}irevi\'{c}}}, \bibinfo {author} {\bibfnamefont {N.}~\bibnamefont
  {Ko\u{s}nik}}, \bibinfo {author} {\bibfnamefont {O.}~\bibnamefont
  {Sumensari}},\ and\ \bibinfo {author} {\bibfnamefont {R.}~\bibnamefont
  {Zukanovich~Funchal}},\ }\href {https://doi.org/10.1007/JHEP11(2016)035}
  {\bibfield  {journal} {\bibinfo  {journal} {JHEP}\ }\textbf {\bibinfo
  {volume} {11}},\ \bibinfo {pages} {035} (\bibinfo {year} {2016})},\ \Eprint
  {https://arxiv.org/abs/1608.07583} {arXiv:1608.07583 [hep-ph]} \BibitemShut
  {NoStop}%
\bibitem [{\citenamefont {Popov}\ and\ \citenamefont
  {White}(2017)}]{Popov:2016fzr}%
  \BibitemOpen
  \bibfield  {author} {\bibinfo {author} {\bibfnamefont {O.}~\bibnamefont
  {Popov}}\ and\ \bibinfo {author} {\bibfnamefont {G.~A.}\ \bibnamefont
  {White}},\ }\href {https://doi.org/10.1016/j.nuclphysb.2017.08.007}
  {\bibfield  {journal} {\bibinfo  {journal} {Nucl. Phys.}\ }\textbf {\bibinfo
  {volume} {B923}},\ \bibinfo {pages} {324} (\bibinfo {year} {2017})},\ \Eprint
  {https://arxiv.org/abs/1611.04566} {arXiv:1611.04566 [hep-ph]} \BibitemShut
  {NoStop}%
\bibitem [{\citenamefont {Coluccio~Leskow}\ \emph {et~al.}(2017)\citenamefont
  {Coluccio~Leskow}, \citenamefont {D'Ambrosio}, \citenamefont {Crivellin},\
  and\ \citenamefont {Müller}}]{ColuccioLeskow:2016dox}%
  \BibitemOpen
  \bibfield  {author} {\bibinfo {author} {\bibfnamefont {E.}~\bibnamefont
  {Coluccio~Leskow}}, \bibinfo {author} {\bibfnamefont {G.}~\bibnamefont
  {D'Ambrosio}}, \bibinfo {author} {\bibfnamefont {A.}~\bibnamefont
  {Crivellin}},\ and\ \bibinfo {author} {\bibfnamefont {D.}~\bibnamefont
  {Müller}},\ }\href {https://doi.org/10.1103/PhysRevD.95.055018} {\bibfield
  {journal} {\bibinfo  {journal} {Phys. Rev.}\ }\textbf {\bibinfo {volume}
  {D95}},\ \bibinfo {pages} {055018} (\bibinfo {year} {2017})},\ \Eprint
  {https://arxiv.org/abs/1612.06858} {arXiv:1612.06858 [hep-ph]} \BibitemShut
  {NoStop}%
\bibitem [{\citenamefont {Buttazzo}\ \emph {et~al.}(2017)\citenamefont
  {Buttazzo}, \citenamefont {Greljo}, \citenamefont {Isidori},\ and\
  \citenamefont {Marzocca}}]{Buttazzo:2017ixm}%
  \BibitemOpen
  \bibfield  {author} {\bibinfo {author} {\bibfnamefont {D.}~\bibnamefont
  {Buttazzo}}, \bibinfo {author} {\bibfnamefont {A.}~\bibnamefont {Greljo}},
  \bibinfo {author} {\bibfnamefont {G.}~\bibnamefont {Isidori}},\ and\ \bibinfo
  {author} {\bibfnamefont {D.}~\bibnamefont {Marzocca}},\ }\href
  {https://doi.org/10.1007/JHEP11(2017)044} {\bibfield  {journal} {\bibinfo
  {journal} {JHEP}\ }\textbf {\bibinfo {volume} {11}},\ \bibinfo {pages} {044}
  (\bibinfo {year} {2017})},\ \Eprint {https://arxiv.org/abs/1706.07808}
  {arXiv:1706.07808 [hep-ph]} \BibitemShut {NoStop}%
\bibitem [{\citenamefont {Angelescu}\ \emph {et~al.}(2018)\citenamefont
  {Angelescu}, \citenamefont {Be\u{c}irevi\'{c}}, \citenamefont {Faroughy},\
  and\ \citenamefont {Sumensari}}]{Angelescu:2018tyl}%
  \BibitemOpen
  \bibfield  {author} {\bibinfo {author} {\bibfnamefont {A.}~\bibnamefont
  {Angelescu}}, \bibinfo {author} {\bibfnamefont {D.}~\bibnamefont
  {Be\u{c}irevi\'{c}}}, \bibinfo {author} {\bibfnamefont {D.~A.}\ \bibnamefont
  {Faroughy}},\ and\ \bibinfo {author} {\bibfnamefont {O.}~\bibnamefont
  {Sumensari}},\ }\href {https://doi.org/10.1007/JHEP10(2018)183} {\bibfield
  {journal} {\bibinfo  {journal} {JHEP}\ }\textbf {\bibinfo {volume} {10}},\
  \bibinfo {pages} {183} (\bibinfo {year} {2018})},\ \Eprint
  {https://arxiv.org/abs/1808.08179} {arXiv:1808.08179 [hep-ph]} \BibitemShut
  {NoStop}%
\bibitem [{\citenamefont {Dor\v{s}ner}, \citenamefont {Fajfer},\ and\
  \citenamefont {Sumensari}(2020)}]{Dorsner:2019itg}%
  \BibitemOpen
  \bibfield  {author} {\bibinfo {author} {\bibfnamefont {I.}~\bibnamefont
  {Dor\v{s}ner}}, \bibinfo {author} {\bibfnamefont {S.}~\bibnamefont
  {Fajfer}},\ and\ \bibinfo {author} {\bibfnamefont {O.}~\bibnamefont
  {Sumensari}},\ }\href {https://doi.org/10.1007/JHEP06(2020)089} {\bibfield
  {journal} {\bibinfo  {journal} {JHEP}\ }\textbf {\bibinfo {volume} {06}},\
  \bibinfo {pages} {089} (\bibinfo {year} {2020})},\ \Eprint
  {https://arxiv.org/abs/1910.03877} {arXiv:1910.03877 [hep-ph]} \BibitemShut
  {NoStop}%
\bibitem [{\citenamefont {Bigaran}, \citenamefont {Gargalionis},\ and\
  \citenamefont {Volkas}(2019)}]{Bigaran:2019bqv}%
  \BibitemOpen
  \bibfield  {author} {\bibinfo {author} {\bibfnamefont {I.}~\bibnamefont
  {Bigaran}}, \bibinfo {author} {\bibfnamefont {J.}~\bibnamefont
  {Gargalionis}},\ and\ \bibinfo {author} {\bibfnamefont {R.~R.}\ \bibnamefont
  {Volkas}},\ }\href {https://doi.org/10.1007/JHEP10(2019)106} {\bibfield
  {journal} {\bibinfo  {journal} {JHEP}\ }\textbf {\bibinfo {volume} {10}},\
  \bibinfo {pages} {106} (\bibinfo {year} {2019})},\ \Eprint
  {https://arxiv.org/abs/1906.01870} {arXiv:1906.01870 [hep-ph]} \BibitemShut
  {NoStop}%
\bibitem [{\citenamefont {Crivellin}\ and\ \citenamefont
  {Saturnino}(2019)}]{Crivellin:2019qnh}%
  \BibitemOpen
  \bibfield  {author} {\bibinfo {author} {\bibfnamefont {A.}~\bibnamefont
  {Crivellin}}\ and\ \bibinfo {author} {\bibfnamefont {F.}~\bibnamefont
  {Saturnino}},\ }\href {https://doi.org/10.1103/PhysRevD.100.115014}
  {\bibfield  {journal} {\bibinfo  {journal} {Phys. Rev. D}\ }\textbf {\bibinfo
  {volume} {100}},\ \bibinfo {pages} {115014} (\bibinfo {year} {2019})},\
  \Eprint {https://arxiv.org/abs/1905.08257} {arXiv:1905.08257 [hep-ph]}
  \BibitemShut {NoStop}%
\bibitem [{\citenamefont {Bordone}\ \emph {et~al.}(2021)\citenamefont
  {Bordone}, \citenamefont {Cat\`a}, \citenamefont {Feldmann},\ and\
  \citenamefont {Mandal}}]{Bordone:2020lnb}%
  \BibitemOpen
  \bibfield  {author} {\bibinfo {author} {\bibfnamefont {M.}~\bibnamefont
  {Bordone}}, \bibinfo {author} {\bibfnamefont {O.}~\bibnamefont {Cat\`a}},
  \bibinfo {author} {\bibfnamefont {T.}~\bibnamefont {Feldmann}},\ and\
  \bibinfo {author} {\bibfnamefont {R.}~\bibnamefont {Mandal}},\ }\href
  {https://doi.org/10.1007/JHEP03(2021)122} {\bibfield  {journal} {\bibinfo
  {journal} {JHEP}\ }\textbf {\bibinfo {volume} {03}},\ \bibinfo {pages} {122}
  (\bibinfo {year} {2021})},\ \Eprint {https://arxiv.org/abs/2010.03297}
  {arXiv:2010.03297 [hep-ph]} \BibitemShut {NoStop}%
\bibitem [{\citenamefont {Crivellin}\ \emph {et~al.}(2021)\citenamefont
  {Crivellin}, \citenamefont {Greub}, \citenamefont {M\"uller},\ and\
  \citenamefont {Saturnino}}]{Crivellin:2020mjs}%
  \BibitemOpen
  \bibfield  {author} {\bibinfo {author} {\bibfnamefont {A.}~\bibnamefont
  {Crivellin}}, \bibinfo {author} {\bibfnamefont {C.}~\bibnamefont {Greub}},
  \bibinfo {author} {\bibfnamefont {D.}~\bibnamefont {M\"uller}},\ and\
  \bibinfo {author} {\bibfnamefont {F.}~\bibnamefont {Saturnino}},\ }\href
  {https://doi.org/10.1007/JHEP02(2021)182} {\bibfield  {journal} {\bibinfo
  {journal} {JHEP}\ }\textbf {\bibinfo {volume} {02}},\ \bibinfo {pages} {182}
  (\bibinfo {year} {2021})},\ \Eprint {https://arxiv.org/abs/2010.06593}
  {arXiv:2010.06593 [hep-ph]} \BibitemShut {NoStop}%
\bibitem [{\citenamefont {Gherardi}, \citenamefont {Marzocca},\ and\
  \citenamefont {Venturini}(2021)}]{Gherardi:2020qhc}%
  \BibitemOpen
  \bibfield  {author} {\bibinfo {author} {\bibfnamefont {V.}~\bibnamefont
  {Gherardi}}, \bibinfo {author} {\bibfnamefont {D.}~\bibnamefont {Marzocca}},\
  and\ \bibinfo {author} {\bibfnamefont {E.}~\bibnamefont {Venturini}},\ }\href
  {https://doi.org/10.1007/JHEP01(2021)138} {\bibfield  {journal} {\bibinfo
  {journal} {JHEP}\ }\textbf {\bibinfo {volume} {01}},\ \bibinfo {pages} {138}
  (\bibinfo {year} {2021})},\ \Eprint {https://arxiv.org/abs/2008.09548}
  {arXiv:2008.09548 [hep-ph]} \BibitemShut {NoStop}%
\bibitem [{\citenamefont {Dash}\ \emph {et~al.}(2021)\citenamefont {Dash},
  \citenamefont {Mishra}, \citenamefont {Patra},\ and\ \citenamefont
  {Sahu}}]{Dash:2021suj}%
  \BibitemOpen
  \bibfield  {author} {\bibinfo {author} {\bibfnamefont {C.}~\bibnamefont
  {Dash}}, \bibinfo {author} {\bibfnamefont {S.}~\bibnamefont {Mishra}},
  \bibinfo {author} {\bibfnamefont {S.}~\bibnamefont {Patra}},\ and\ \bibinfo
  {author} {\bibfnamefont {P.}~\bibnamefont {Sahu}},\ }\href@noop {} {\
  (\bibinfo {year} {2021})},\ \Eprint {https://arxiv.org/abs/2109.12536}
  {arXiv:2109.12536 [hep-ph]} \BibitemShut {NoStop}%
\bibitem [{\citenamefont {Greljo}, \citenamefont {Stangl},\ and\ \citenamefont
  {Thomsen}(2021)}]{Greljo:2021xmg}%
  \BibitemOpen
  \bibfield  {author} {\bibinfo {author} {\bibfnamefont {A.}~\bibnamefont
  {Greljo}}, \bibinfo {author} {\bibfnamefont {P.}~\bibnamefont {Stangl}},\
  and\ \bibinfo {author} {\bibfnamefont {A.~E.}\ \bibnamefont {Thomsen}},\
  }\href {https://doi.org/10.1016/j.physletb.2021.136554} {\bibfield  {journal}
  {\bibinfo  {journal} {Phys. Lett. B}\ }\textbf {\bibinfo {volume} {820}},\
  \bibinfo {pages} {136554} (\bibinfo {year} {2021})},\ \Eprint
  {https://arxiv.org/abs/2103.13991} {arXiv:2103.13991 [hep-ph]} \BibitemShut
  {NoStop}%
\bibitem [{\citenamefont {Marzocca}, \citenamefont {Trifinopoulos},\ and\
  \citenamefont {Venturini}(2021)}]{Marzocca:2021miv}%
  \BibitemOpen
  \bibfield  {author} {\bibinfo {author} {\bibfnamefont {D.}~\bibnamefont
  {Marzocca}}, \bibinfo {author} {\bibfnamefont {S.}~\bibnamefont
  {Trifinopoulos}},\ and\ \bibinfo {author} {\bibfnamefont {E.}~\bibnamefont
  {Venturini}},\ }\href@noop {} {\  (\bibinfo {year} {2021})},\ \Eprint
  {https://arxiv.org/abs/2106.15630} {arXiv:2106.15630 [hep-ph]} \BibitemShut
  {NoStop}%
\bibitem [{\citenamefont {Marzocca}\ and\ \citenamefont
  {Trifinopoulos}(2021)}]{Marzocca:2021azj}%
  \BibitemOpen
  \bibfield  {author} {\bibinfo {author} {\bibfnamefont {D.}~\bibnamefont
  {Marzocca}}\ and\ \bibinfo {author} {\bibfnamefont {S.}~\bibnamefont
  {Trifinopoulos}},\ }\href {https://doi.org/10.1103/PhysRevLett.127.061803}
  {\bibfield  {journal} {\bibinfo  {journal} {Phys. Rev. Lett.}\ }\textbf
  {\bibinfo {volume} {127}},\ \bibinfo {pages} {061803} (\bibinfo {year}
  {2021})},\ \Eprint {https://arxiv.org/abs/2104.05730} {arXiv:2104.05730
  [hep-ph]} \BibitemShut {NoStop}%
\bibitem [{\citenamefont {Mohapatra}, \citenamefont {Rajeev},\ and\
  \citenamefont {Dutta}(2021)}]{Mohapatra:2021ynn}%
  \BibitemOpen
  \bibfield  {author} {\bibinfo {author} {\bibfnamefont {M.~K.}\ \bibnamefont
  {Mohapatra}}, \bibinfo {author} {\bibfnamefont {N.}~\bibnamefont {Rajeev}},\
  and\ \bibinfo {author} {\bibfnamefont {R.}~\bibnamefont {Dutta}},\
  }\href@noop {} {\  (\bibinfo {year} {2021})},\ \Eprint
  {https://arxiv.org/abs/2108.10106} {arXiv:2108.10106 [hep-ph]} \BibitemShut
  {NoStop}%
\bibitem [{\citenamefont {Murgui}\ and\ \citenamefont
  {Wise}(2021)}]{Murgui:2021bdy}%
  \BibitemOpen
  \bibfield  {author} {\bibinfo {author} {\bibfnamefont {C.}~\bibnamefont
  {Murgui}}\ and\ \bibinfo {author} {\bibfnamefont {M.~B.}\ \bibnamefont
  {Wise}},\ }\href {https://doi.org/10.1103/PhysRevD.104.035017} {\bibfield
  {journal} {\bibinfo  {journal} {Phys. Rev. D}\ }\textbf {\bibinfo {volume}
  {104}},\ \bibinfo {pages} {035017} (\bibinfo {year} {2021})},\ \Eprint
  {https://arxiv.org/abs/2105.14029} {arXiv:2105.14029 [hep-ph]} \BibitemShut
  {NoStop}%
\bibitem [{\citenamefont {Perez}, \citenamefont {Murgui},\ and\ \citenamefont
  {Plascencia}(2021)}]{Perez:2021ddi}%
  \BibitemOpen
  \bibfield  {author} {\bibinfo {author} {\bibfnamefont {P.~F.}\ \bibnamefont
  {Perez}}, \bibinfo {author} {\bibfnamefont {C.}~\bibnamefont {Murgui}},\ and\
  \bibinfo {author} {\bibfnamefont {A.~D.}\ \bibnamefont {Plascencia}},\ }\href
  {https://doi.org/10.1103/PhysRevD.104.035041} {\bibfield  {journal} {\bibinfo
   {journal} {Phys. Rev. D}\ }\textbf {\bibinfo {volume} {104}},\ \bibinfo
  {pages} {035041} (\bibinfo {year} {2021})},\ \Eprint
  {https://arxiv.org/abs/2104.11229} {arXiv:2104.11229 [hep-ph]} \BibitemShut
  {NoStop}%
\bibitem [{\citenamefont {Singirala}, \citenamefont {Sahoo},\ and\
  \citenamefont {Mohanta}(2021)}]{Singirala:2021gok}%
  \BibitemOpen
  \bibfield  {author} {\bibinfo {author} {\bibfnamefont {S.}~\bibnamefont
  {Singirala}}, \bibinfo {author} {\bibfnamefont {S.}~\bibnamefont {Sahoo}},\
  and\ \bibinfo {author} {\bibfnamefont {R.}~\bibnamefont {Mohanta}},\
  }\href@noop {} {\  (\bibinfo {year} {2021})},\ \Eprint
  {https://arxiv.org/abs/2106.03735} {arXiv:2106.03735 [hep-ph]} \BibitemShut
  {NoStop}%
\bibitem [{\citenamefont {Zhang}(2021)}]{Zhang:2021dgl}%
  \BibitemOpen
  \bibfield  {author} {\bibinfo {author} {\bibfnamefont {D.}~\bibnamefont
  {Zhang}},\ }\href {https://doi.org/10.1007/JHEP07(2021)069} {\bibfield
  {journal} {\bibinfo  {journal} {JHEP}\ }\textbf {\bibinfo {volume} {07}},\
  \bibinfo {pages} {069} (\bibinfo {year} {2021})},\ \Eprint
  {https://arxiv.org/abs/2105.08670} {arXiv:2105.08670 [hep-ph]} \BibitemShut
  {NoStop}%
\bibitem [{\citenamefont {Angelescu}\ \emph {et~al.}(2021)\citenamefont
  {Angelescu}, \citenamefont {Be\v{c}irevi\'c}, \citenamefont {Faroughy},
  \citenamefont {Jaffredo},\ and\ \citenamefont
  {Sumensari}}]{Angelescu:2021lln}%
  \BibitemOpen
  \bibfield  {author} {\bibinfo {author} {\bibfnamefont {A.}~\bibnamefont
  {Angelescu}}, \bibinfo {author} {\bibfnamefont {D.}~\bibnamefont
  {Be\v{c}irevi\'c}}, \bibinfo {author} {\bibfnamefont {D.~A.}\ \bibnamefont
  {Faroughy}}, \bibinfo {author} {\bibfnamefont {F.}~\bibnamefont {Jaffredo}},\
  and\ \bibinfo {author} {\bibfnamefont {O.}~\bibnamefont {Sumensari}},\ }\href
  {https://doi.org/10.1103/PhysRevD.104.055017} {\bibfield  {journal} {\bibinfo
   {journal} {Phys. Rev. D}\ }\textbf {\bibinfo {volume} {104}},\ \bibinfo
  {pages} {055017} (\bibinfo {year} {2021})},\ \Eprint
  {https://arxiv.org/abs/2103.12504} {arXiv:2103.12504 [hep-ph]} \BibitemShut
  {NoStop}%
\bibitem [{\citenamefont {Dor\v{s}ner}, \citenamefont {Fajfer},\ and\
  \citenamefont {Saad}(2020)}]{Dorsner:2020aaz}%
  \BibitemOpen
  \bibfield  {author} {\bibinfo {author} {\bibfnamefont {I.}~\bibnamefont
  {Dor\v{s}ner}}, \bibinfo {author} {\bibfnamefont {S.}~\bibnamefont
  {Fajfer}},\ and\ \bibinfo {author} {\bibfnamefont {S.}~\bibnamefont {Saad}},\
  }\href {https://doi.org/10.1103/PhysRevD.102.075007} {\bibfield  {journal}
  {\bibinfo  {journal} {Phys. Rev. D}\ }\textbf {\bibinfo {volume} {102}},\
  \bibinfo {pages} {075007} (\bibinfo {year} {2020})},\ \Eprint
  {https://arxiv.org/abs/2006.11624} {arXiv:2006.11624 [hep-ph]} \BibitemShut
  {NoStop}%
\bibitem [{\citenamefont {Kowalska}, \citenamefont {Sessolo},\ and\
  \citenamefont {Yamamoto}(2019)}]{Kowalska:2018ulj}%
  \BibitemOpen
  \bibfield  {author} {\bibinfo {author} {\bibfnamefont {K.}~\bibnamefont
  {Kowalska}}, \bibinfo {author} {\bibfnamefont {E.~M.}\ \bibnamefont
  {Sessolo}},\ and\ \bibinfo {author} {\bibfnamefont {Y.}~\bibnamefont
  {Yamamoto}},\ }\href {https://doi.org/10.1103/PhysRevD.99.055007} {\bibfield
  {journal} {\bibinfo  {journal} {Phys. Rev.}\ }\textbf {\bibinfo {volume}
  {D99}},\ \bibinfo {pages} {055007} (\bibinfo {year} {2019})},\ \Eprint
  {https://arxiv.org/abs/1812.06851} {arXiv:1812.06851 [hep-ph]} \BibitemShut
  {NoStop}%
\bibitem [{\citenamefont {Aebischer}\ \emph {et~al.}(2021)\citenamefont
  {Aebischer}, \citenamefont {Dekens}, \citenamefont {Jenkins}, \citenamefont
  {Manohar}, \citenamefont {Sengupta},\ and\ \citenamefont
  {Stoffer}}]{Aebischer:2021uvt}%
  \BibitemOpen
  \bibfield  {author} {\bibinfo {author} {\bibfnamefont {J.}~\bibnamefont
  {Aebischer}}, \bibinfo {author} {\bibfnamefont {W.}~\bibnamefont {Dekens}},
  \bibinfo {author} {\bibfnamefont {E.~E.}\ \bibnamefont {Jenkins}}, \bibinfo
  {author} {\bibfnamefont {A.~V.}\ \bibnamefont {Manohar}}, \bibinfo {author}
  {\bibfnamefont {D.}~\bibnamefont {Sengupta}},\ and\ \bibinfo {author}
  {\bibfnamefont {P.}~\bibnamefont {Stoffer}},\ }\href
  {https://doi.org/10.1007/JHEP07(2021)107} {\bibfield  {journal} {\bibinfo
  {journal} {JHEP}\ }\textbf {\bibinfo {volume} {07}},\ \bibinfo {pages} {107}
  (\bibinfo {year} {2021})},\ \Eprint {https://arxiv.org/abs/2102.08954}
  {arXiv:2102.08954 [hep-ph]} \BibitemShut {NoStop}%
\bibitem [{\citenamefont {Crivellin}\ and\ \citenamefont
  {Hoferichter}(2020)}]{Crivellin:2019mvj}%
  \BibitemOpen
  \bibfield  {author} {\bibinfo {author} {\bibfnamefont {A.}~\bibnamefont
  {Crivellin}}\ and\ \bibinfo {author} {\bibfnamefont {M.}~\bibnamefont
  {Hoferichter}},\ }\href {https://doi.org/10.22323/1.360.0009} {\bibfield
  {journal} {\bibinfo  {journal} {PoS}\ }\textbf {\bibinfo {volume}
  {ALPS2019}},\ \bibinfo {pages} {009} (\bibinfo {year} {2020})},\ \Eprint
  {https://arxiv.org/abs/1905.03789} {arXiv:1905.03789 [hep-ph]} \BibitemShut
  {NoStop}%
\bibitem [{\citenamefont {Baker}, \citenamefont {Cox},\ and\ \citenamefont
  {Volkas}(2021)}]{Baker:2021yli}%
  \BibitemOpen
  \bibfield  {author} {\bibinfo {author} {\bibfnamefont {M.~J.}\ \bibnamefont
  {Baker}}, \bibinfo {author} {\bibfnamefont {P.}~\bibnamefont {Cox}},\ and\
  \bibinfo {author} {\bibfnamefont {R.~R.}\ \bibnamefont {Volkas}},\ }\href
  {https://doi.org/10.1007/JHEP05(2021)174} {\bibfield  {journal} {\bibinfo
  {journal} {JHEP}\ }\textbf {\bibinfo {volume} {05}},\ \bibinfo {pages} {174}
  (\bibinfo {year} {2021})},\ \Eprint {https://arxiv.org/abs/2103.13401}
  {arXiv:2103.13401 [hep-ph]} \BibitemShut {NoStop}%
\bibitem [{\citenamefont {Athron}\ \emph {et~al.}(2021)\citenamefont {Athron},
  \citenamefont {Bal\'azs}, \citenamefont {Jacob}, \citenamefont {Kotlarski},
  \citenamefont {St\"ockinger},\ and\ \citenamefont
  {St\"ockinger-Kim}}]{Athron:2021iuf}%
  \BibitemOpen
  \bibfield  {author} {\bibinfo {author} {\bibfnamefont {P.}~\bibnamefont
  {Athron}}, \bibinfo {author} {\bibfnamefont {C.}~\bibnamefont {Bal\'azs}},
  \bibinfo {author} {\bibfnamefont {D.~H.}\ \bibnamefont {Jacob}}, \bibinfo
  {author} {\bibfnamefont {W.}~\bibnamefont {Kotlarski}}, \bibinfo {author}
  {\bibfnamefont {D.}~\bibnamefont {St\"ockinger}},\ and\ \bibinfo {author}
  {\bibfnamefont {H.}~\bibnamefont {St\"ockinger-Kim}},\ }\href@noop {} {\
  (\bibinfo {year} {2021})},\ \Eprint {https://arxiv.org/abs/2104.03691}
  {arXiv:2104.03691 [hep-ph]} \BibitemShut {NoStop}%
\bibitem [{\citenamefont {Xing}, \citenamefont {Zhang},\ and\ \citenamefont
  {Zhou}(2008)}]{Xing:2007fb}%
  \BibitemOpen
  \bibfield  {author} {\bibinfo {author} {\bibfnamefont {Z.-z.}\ \bibnamefont
  {Xing}}, \bibinfo {author} {\bibfnamefont {H.}~\bibnamefont {Zhang}},\ and\
  \bibinfo {author} {\bibfnamefont {S.}~\bibnamefont {Zhou}},\ }\href
  {https://doi.org/10.1103/PhysRevD.77.113016} {\bibfield  {journal} {\bibinfo
  {journal} {Phys. Rev. D}\ }\textbf {\bibinfo {volume} {77}},\ \bibinfo
  {pages} {113016} (\bibinfo {year} {2008})},\ \Eprint
  {https://arxiv.org/abs/0712.1419} {arXiv:0712.1419 [hep-ph]} \BibitemShut
  {NoStop}%
\bibitem [{\citenamefont {Aad}\ \emph {et~al.}(2020{\natexlab{a}})\citenamefont
  {Aad} \emph {et~al.}}]{Aad:2020iuy}%
  \BibitemOpen
  \bibfield  {author} {\bibinfo {author} {\bibfnamefont {G.}~\bibnamefont
  {Aad}} \emph {et~al.} (\bibinfo {collaboration} {ATLAS}),\ }\href
  {https://doi.org/10.1007/JHEP10(2020)112} {\bibfield  {journal} {\bibinfo
  {journal} {JHEP}\ }\textbf {\bibinfo {volume} {10}},\ \bibinfo {pages} {112}
  (\bibinfo {year} {2020}{\natexlab{a}})},\ \Eprint
  {https://arxiv.org/abs/2006.05872} {arXiv:2006.05872 [hep-ex]} \BibitemShut
  {NoStop}%
\bibitem [{\citenamefont {Kirch}\ and\ \citenamefont
  {Schmidt-Wellenburg}(2020)}]{Kirch:2020lbo}%
  \BibitemOpen
  \bibfield  {author} {\bibinfo {author} {\bibfnamefont {K.}~\bibnamefont
  {Kirch}}\ and\ \bibinfo {author} {\bibfnamefont {P.}~\bibnamefont
  {Schmidt-Wellenburg}},\ }\href {https://doi.org/10.1051/epjconf/202023401007}
  {\bibfield  {journal} {\bibinfo  {journal} {EPJ Web Conf.}\ }\textbf
  {\bibinfo {volume} {234}},\ \bibinfo {pages} {01007} (\bibinfo {year}
  {2020})},\ \Eprint {https://arxiv.org/abs/2003.00717} {arXiv:2003.00717
  [hep-ph]} \BibitemShut {NoStop}%
\bibitem [{\citenamefont {Adelmann}\ \emph {et~al.}(2021)\citenamefont
  {Adelmann} \emph {et~al.}}]{Adelmann:2021udj}%
  \BibitemOpen
  \bibfield  {author} {\bibinfo {author} {\bibfnamefont {A.}~\bibnamefont
  {Adelmann}} \emph {et~al.},\ }\href@noop {} {\  (\bibinfo {year} {2021})},\
  \Eprint {https://arxiv.org/abs/2102.08838} {arXiv:2102.08838 [hep-ex]}
  \BibitemShut {NoStop}%
\bibitem [{\citenamefont {Chislett}(2016)}]{Chislett:2016jau}%
  \BibitemOpen
  \bibfield  {author} {\bibinfo {author} {\bibfnamefont {R.}~\bibnamefont
  {Chislett}} (\bibinfo {collaboration} {Muon g-2}),\ }\href
  {https://doi.org/10.1051/epjconf/201611801005} {\bibfield  {journal}
  {\bibinfo  {journal} {EPJ Web Conf.}\ }\textbf {\bibinfo {volume} {118}},\
  \bibinfo {pages} {01005} (\bibinfo {year} {2016})}\BibitemShut {NoStop}%
\bibitem [{\citenamefont {Greljo}\ and\ \citenamefont
  {Marzocca}(2017)}]{Greljo:2017vvb}%
  \BibitemOpen
  \bibfield  {author} {\bibinfo {author} {\bibfnamefont {A.}~\bibnamefont
  {Greljo}}\ and\ \bibinfo {author} {\bibfnamefont {D.}~\bibnamefont
  {Marzocca}},\ }\href {https://doi.org/10.1140/epjc/s10052-017-5119-8}
  {\bibfield  {journal} {\bibinfo  {journal} {Eur. Phys. J. C}\ }\textbf
  {\bibinfo {volume} {77}},\ \bibinfo {pages} {548} (\bibinfo {year} {2017})},\
  \Eprint {https://arxiv.org/abs/1704.09015} {arXiv:1704.09015 [hep-ph]}
  \BibitemShut {NoStop}%
\bibitem [{\citenamefont {Tanabashi}\ \emph {et~al.}(2018)\citenamefont
  {Tanabashi} \emph {et~al.}}]{PDG}%
  \BibitemOpen
  \bibfield  {author} {\bibinfo {author} {\bibfnamefont {M.}~\bibnamefont
  {Tanabashi}} \emph {et~al.} (\bibinfo {collaboration} {Particle Data
  Group}),\ }\href {https://doi.org/10.1103/PhysRevD.98.030001} {\bibfield
  {journal} {\bibinfo  {journal} {Phys. Rev.}\ }\textbf {\bibinfo {volume}
  {D98}},\ \bibinfo {pages} {030001} (\bibinfo {year} {2018})}\BibitemShut
  {NoStop}%
\bibitem [{\citenamefont {Crivellin}\ and\ \citenamefont
  {Hoferichter}(2021)}]{Crivellin:2021rbq}%
  \BibitemOpen
  \bibfield  {author} {\bibinfo {author} {\bibfnamefont {A.}~\bibnamefont
  {Crivellin}}\ and\ \bibinfo {author} {\bibfnamefont {M.}~\bibnamefont
  {Hoferichter}},\ }\href {https://doi.org/10.1007/JHEP07(2021)135} {\bibfield
  {journal} {\bibinfo  {journal} {JHEP}\ }\textbf {\bibinfo {volume} {07}},\
  \bibinfo {pages} {135} (\bibinfo {year} {2021})},\ \Eprint
  {https://arxiv.org/abs/2104.03202} {arXiv:2104.03202 [hep-ph]} \BibitemShut
  {NoStop}%
\bibitem [{\citenamefont {Arnan}\ \emph {et~al.}(2019)\citenamefont {Arnan},
  \citenamefont {Be\u{c}irevi\'{c}}, \citenamefont {Mescia},\ and\
  \citenamefont {Sumensari}}]{Arnan:2019olv}%
  \BibitemOpen
  \bibfield  {author} {\bibinfo {author} {\bibfnamefont {P.}~\bibnamefont
  {Arnan}}, \bibinfo {author} {\bibfnamefont {D.}~\bibnamefont
  {Be\u{c}irevi\'{c}}}, \bibinfo {author} {\bibfnamefont {F.}~\bibnamefont
  {Mescia}},\ and\ \bibinfo {author} {\bibfnamefont {O.}~\bibnamefont
  {Sumensari}},\ }\href {https://doi.org/10.1007/JHEP02(2019)109} {\bibfield
  {journal} {\bibinfo  {journal} {JHEP}\ }\textbf {\bibinfo {volume} {02}},\
  \bibinfo {pages} {109} (\bibinfo {year} {2019})},\ \Eprint
  {https://arxiv.org/abs/1901.06315} {arXiv:1901.06315 [hep-ph]} \BibitemShut
  {NoStop}%
\bibitem [{\citenamefont {Fajfer}, \citenamefont {Kamenik},\ and\ \citenamefont
  {Tammaro}(2021)}]{Fajfer:2021cxa}%
  \BibitemOpen
  \bibfield  {author} {\bibinfo {author} {\bibfnamefont {S.}~\bibnamefont
  {Fajfer}}, \bibinfo {author} {\bibfnamefont {J.~F.}\ \bibnamefont
  {Kamenik}},\ and\ \bibinfo {author} {\bibfnamefont {M.}~\bibnamefont
  {Tammaro}},\ }\href {https://doi.org/10.1007/JHEP06(2021)099} {\bibfield
  {journal} {\bibinfo  {journal} {JHEP}\ }\textbf {\bibinfo {volume} {06}},\
  \bibinfo {pages} {099} (\bibinfo {year} {2021})},\ \Eprint
  {https://arxiv.org/abs/2103.10859} {arXiv:2103.10859 [hep-ph]} \BibitemShut
  {NoStop}%
\bibitem [{\citenamefont {Feruglio}, \citenamefont {Paradisi},\ and\
  \citenamefont {Sumensari}(2018)}]{Feruglio:2018fxo}%
  \BibitemOpen
  \bibfield  {author} {\bibinfo {author} {\bibfnamefont {F.}~\bibnamefont
  {Feruglio}}, \bibinfo {author} {\bibfnamefont {P.}~\bibnamefont {Paradisi}},\
  and\ \bibinfo {author} {\bibfnamefont {O.}~\bibnamefont {Sumensari}},\ }\href
  {https://doi.org/10.1007/JHEP11(2018)191} {\bibfield  {journal} {\bibinfo
  {journal} {JHEP}\ }\textbf {\bibinfo {volume} {11}},\ \bibinfo {pages} {191}
  (\bibinfo {year} {2018})},\ \Eprint {https://arxiv.org/abs/1806.10155}
  {arXiv:1806.10155 [hep-ph]} \BibitemShut {NoStop}%
\bibitem [{\citenamefont {Sirunyan}\ \emph {et~al.}(2021)\citenamefont
  {Sirunyan} \emph {et~al.}}]{CMS:2020xwi}%
  \BibitemOpen
  \bibfield  {author} {\bibinfo {author} {\bibfnamefont {A.~M.}\ \bibnamefont
  {Sirunyan}} \emph {et~al.} (\bibinfo {collaboration} {CMS}),\ }\href
  {https://doi.org/10.1007/JHEP01(2021)148} {\bibfield  {journal} {\bibinfo
  {journal} {JHEP}\ }\textbf {\bibinfo {volume} {01}},\ \bibinfo {pages} {148}
  (\bibinfo {year} {2021})},\ \Eprint {https://arxiv.org/abs/2009.04363}
  {arXiv:2009.04363 [hep-ex]} \BibitemShut {NoStop}%
\bibitem [{\citenamefont {Aad}\ \emph {et~al.}(2021)\citenamefont {Aad} \emph
  {et~al.}}]{ATLAS:2020fzp}%
  \BibitemOpen
  \bibfield  {author} {\bibinfo {author} {\bibfnamefont {G.}~\bibnamefont
  {Aad}} \emph {et~al.} (\bibinfo {collaboration} {ATLAS}),\ }\href
  {https://doi.org/10.1016/j.physletb.2020.135980} {\bibfield  {journal}
  {\bibinfo  {journal} {Phys. Lett. B}\ }\textbf {\bibinfo {volume} {812}},\
  \bibinfo {pages} {135980} (\bibinfo {year} {2021})},\ \Eprint
  {https://arxiv.org/abs/2007.07830} {arXiv:2007.07830 [hep-ex]} \BibitemShut
  {NoStop}%
\bibitem [{\citenamefont {Crivellin}, \citenamefont {Mueller},\ and\
  \citenamefont {Saturnino}(2021)}]{Crivellin:2020tsz}%
  \BibitemOpen
  \bibfield  {author} {\bibinfo {author} {\bibfnamefont {A.}~\bibnamefont
  {Crivellin}}, \bibinfo {author} {\bibfnamefont {D.}~\bibnamefont {Mueller}},\
  and\ \bibinfo {author} {\bibfnamefont {F.}~\bibnamefont {Saturnino}},\ }\href
  {https://doi.org/10.1103/PhysRevLett.127.021801} {\bibfield  {journal}
  {\bibinfo  {journal} {Phys. Rev. Lett.}\ }\textbf {\bibinfo {volume} {127}},\
  \bibinfo {pages} {021801} (\bibinfo {year} {2021})},\ \Eprint
  {https://arxiv.org/abs/2008.02643} {arXiv:2008.02643 [hep-ph]} \BibitemShut
  {NoStop}%
\bibitem [{\citenamefont {Aad}\ \emph {et~al.}(2020{\natexlab{b}})\citenamefont
  {Aad} \emph {et~al.}}]{ATLAS:2019old}%
  \BibitemOpen
  \bibfield  {author} {\bibinfo {author} {\bibfnamefont {G.}~\bibnamefont
  {Aad}} \emph {et~al.} (\bibinfo {collaboration} {ATLAS}),\ }\href
  {https://doi.org/10.1016/j.physletb.2019.135148} {\bibfield  {journal}
  {\bibinfo  {journal} {Phys. Lett. B}\ }\textbf {\bibinfo {volume} {801}},\
  \bibinfo {pages} {135148} (\bibinfo {year} {2020}{\natexlab{b}})},\ \Eprint
  {https://arxiv.org/abs/1909.10235} {arXiv:1909.10235 [hep-ex]} \BibitemShut
  {NoStop}%
\bibitem [{\citenamefont {Kumar}(2016)}]{Kumar:2016omp}%
  \BibitemOpen
  \bibfield  {author} {\bibinfo {author} {\bibfnamefont {G.}~\bibnamefont
  {Kumar}},\ }\href {https://doi.org/10.1103/PhysRevD.94.014022} {\bibfield
  {journal} {\bibinfo  {journal} {Phys. Rev. D}\ }\textbf {\bibinfo {volume}
  {94}},\ \bibinfo {pages} {014022} (\bibinfo {year} {2016})},\ \Eprint
  {https://arxiv.org/abs/1603.00346} {arXiv:1603.00346 [hep-ph]} \BibitemShut
  {NoStop}%
\bibitem [{\citenamefont {Bobeth}\ and\ \citenamefont
  {Buras}(2018)}]{Bobeth:2017ecx}%
  \BibitemOpen
  \bibfield  {author} {\bibinfo {author} {\bibfnamefont {C.}~\bibnamefont
  {Bobeth}}\ and\ \bibinfo {author} {\bibfnamefont {A.~J.}\ \bibnamefont
  {Buras}},\ }\href {https://doi.org/10.1007/JHEP02(2018)101} {\bibfield
  {journal} {\bibinfo  {journal} {JHEP}\ }\textbf {\bibinfo {volume} {02}},\
  \bibinfo {pages} {101} (\bibinfo {year} {2018})},\ \Eprint
  {https://arxiv.org/abs/1712.01295} {arXiv:1712.01295 [hep-ph]} \BibitemShut
  {NoStop}%
\bibitem [{\citenamefont {Mandal}\ and\ \citenamefont
  {Pich}(2019)}]{Mandal:2019gff}%
  \BibitemOpen
  \bibfield  {author} {\bibinfo {author} {\bibfnamefont {R.}~\bibnamefont
  {Mandal}}\ and\ \bibinfo {author} {\bibfnamefont {A.}~\bibnamefont {Pich}},\
  }\href {https://doi.org/10.1007/JHEP12(2019)089} {\bibfield  {journal}
  {\bibinfo  {journal} {JHEP}\ }\textbf {\bibinfo {volume} {12}},\ \bibinfo
  {pages} {089} (\bibinfo {year} {2019})},\ \Eprint
  {https://arxiv.org/abs/1908.11155} {arXiv:1908.11155 [hep-ph]} \BibitemShut
  {NoStop}%
\bibitem [{\citenamefont {Grossman}\ and\ \citenamefont
  {Nir}(1997)}]{Grossman:1997sk}%
  \BibitemOpen
  \bibfield  {author} {\bibinfo {author} {\bibfnamefont {Y.}~\bibnamefont
  {Grossman}}\ and\ \bibinfo {author} {\bibfnamefont {Y.}~\bibnamefont {Nir}},\
  }\href {https://doi.org/10.1016/S0370-2693(97)00210-4} {\bibfield  {journal}
  {\bibinfo  {journal} {Phys. Lett. B}\ }\textbf {\bibinfo {volume} {398}},\
  \bibinfo {pages} {163} (\bibinfo {year} {1997})},\ \Eprint
  {https://arxiv.org/abs/hep-ph/9701313} {arXiv:hep-ph/9701313} \BibitemShut
  {NoStop}%
\bibitem [{\citenamefont {He}, \citenamefont {Valencia},\ and\ \citenamefont
  {Wong}(2018)}]{He:2018uey}%
  \BibitemOpen
  \bibfield  {author} {\bibinfo {author} {\bibfnamefont {X.-G.}\ \bibnamefont
  {He}}, \bibinfo {author} {\bibfnamefont {G.}~\bibnamefont {Valencia}},\ and\
  \bibinfo {author} {\bibfnamefont {K.}~\bibnamefont {Wong}},\ }\href
  {https://doi.org/10.1140/epjc/s10052-018-5964-0} {\bibfield  {journal}
  {\bibinfo  {journal} {Eur. Phys. J. C}\ }\textbf {\bibinfo {volume} {78}},\
  \bibinfo {pages} {472} (\bibinfo {year} {2018})},\ \bibinfo {note} {[Erratum:
  Eur.Phys.J.C 80, 738 (2020)]},\ \Eprint {https://arxiv.org/abs/1804.07449}
  {arXiv:1804.07449 [hep-ph]} \BibitemShut {NoStop}%
\bibitem [{\citenamefont {Ahn}\ \emph {et~al.}(2020)\citenamefont {Ahn} \emph
  {et~al.}}]{KOTO:2020prk}%
  \BibitemOpen
  \bibfield  {author} {\bibinfo {author} {\bibfnamefont {J.~K.}\ \bibnamefont
  {Ahn}} \emph {et~al.} (\bibinfo {collaboration} {KOTO}),\ }\href
  {https://doi.org/10.1103/PhysRevLett.126.121801} {\bibfield  {journal}
  {\bibinfo  {journal} {Phys. Rev. Lett.}\ }\textbf {\bibinfo {volume} {126}},\
  \bibinfo {pages} {121801} (\bibinfo {year} {2020})},\ \Eprint
  {https://arxiv.org/abs/2012.07571} {arXiv:2012.07571 [hep-ex]} \BibitemShut
  {NoStop}%
\bibitem [{\citenamefont {Cortina~Gil}\ \emph {et~al.}(2021)\citenamefont
  {Cortina~Gil} \emph {et~al.}}]{NA62:2021zjw}%
  \BibitemOpen
  \bibfield  {author} {\bibinfo {author} {\bibfnamefont {E.}~\bibnamefont
  {Cortina~Gil}} \emph {et~al.} (\bibinfo {collaboration} {NA62}),\ }\href
  {https://doi.org/10.1007/JHEP06(2021)093} {\bibfield  {journal} {\bibinfo
  {journal} {JHEP}\ }\textbf {\bibinfo {volume} {06}},\ \bibinfo {pages} {093}
  (\bibinfo {year} {2021})},\ \Eprint {https://arxiv.org/abs/2103.15389}
  {arXiv:2103.15389 [hep-ex]} \BibitemShut {NoStop}%
\bibitem [{\citenamefont {Aaboud}\ \emph {et~al.}(2017)\citenamefont {Aaboud}
  \emph {et~al.}}]{Aaboud:2017buh}%
  \BibitemOpen
  \bibfield  {author} {\bibinfo {author} {\bibfnamefont {M.}~\bibnamefont
  {Aaboud}} \emph {et~al.} (\bibinfo {collaboration} {ATLAS}),\ }\href
  {https://doi.org/10.1007/JHEP10(2017)182} {\bibfield  {journal} {\bibinfo
  {journal} {JHEP}\ }\textbf {\bibinfo {volume} {10}},\ \bibinfo {pages} {182}
  (\bibinfo {year} {2017})},\ \Eprint {https://arxiv.org/abs/1707.02424}
  {arXiv:1707.02424 [hep-ex]} \BibitemShut {NoStop}%
\bibitem [{\citenamefont {Fuentes-Martin}\ \emph {et~al.}(2020)\citenamefont
  {Fuentes-Martin}, \citenamefont {Greljo}, \citenamefont {Martin~Camalich},\
  and\ \citenamefont {Ruiz-Alvarez}}]{Fuentes-Martin:2020lea}%
  \BibitemOpen
  \bibfield  {author} {\bibinfo {author} {\bibfnamefont {J.}~\bibnamefont
  {Fuentes-Martin}}, \bibinfo {author} {\bibfnamefont {A.}~\bibnamefont
  {Greljo}}, \bibinfo {author} {\bibfnamefont {J.}~\bibnamefont
  {Martin~Camalich}},\ and\ \bibinfo {author} {\bibfnamefont {J.~D.}\
  \bibnamefont {Ruiz-Alvarez}},\ }\href
  {https://doi.org/10.1007/JHEP11(2020)080} {\bibfield  {journal} {\bibinfo
  {journal} {JHEP}\ }\textbf {\bibinfo {volume} {11}},\ \bibinfo {pages} {080}
  (\bibinfo {year} {2020})},\ \Eprint {https://arxiv.org/abs/2003.12421}
  {arXiv:2003.12421 [hep-ph]} \BibitemShut {NoStop}%
\bibitem [{\citenamefont {González-Sprinberg}\ and\ \citenamefont
  {Vidal}(2017)}]{taummrev}%
  \BibitemOpen
  \bibfield  {author} {\bibinfo {author} {\bibfnamefont {G.}~\bibnamefont
  {González-Sprinberg}}\ and\ \bibinfo {author} {\bibfnamefont
  {J.}~\bibnamefont {Vidal}},\ }\href
  {https://doi.org/10.1088/1742-6596/912/1/012001} {\bibfield  {journal}
  {\bibinfo  {journal} {Journal of Physics: Conference Series}\ }\textbf
  {\bibinfo {volume} {912}},\ \bibinfo {pages} {012001} (\bibinfo {year}
  {2017})}\BibitemShut {NoStop}%
\bibitem [{\citenamefont {Altmannshofer}\ \emph {et~al.}(2019)\citenamefont
  {Altmannshofer} \emph {et~al.}}]{Belle-II:2018jsg}%
  \BibitemOpen
  \bibfield  {author} {\bibinfo {author} {\bibfnamefont {W.}~\bibnamefont
  {Altmannshofer}} \emph {et~al.} (\bibinfo {collaboration} {Belle-II}),\
  }\href {https://doi.org/10.1093/ptep/ptz106} {\bibfield  {journal} {\bibinfo
  {journal} {PTEP}\ }\textbf {\bibinfo {volume} {2019}},\ \bibinfo {pages}
  {123C01} (\bibinfo {year} {2019})},\ \bibinfo {note} {[Erratum: PTEP 2020,
  029201 (2020)]},\ \Eprint {https://arxiv.org/abs/1808.10567}
  {arXiv:1808.10567 [hep-ex]} \BibitemShut {NoStop}%
\bibitem [{\citenamefont {Froggatt}\ and\ \citenamefont
  {Nielsen}(1979)}]{FROGGATT1979277}%
  \BibitemOpen
  \bibfield  {author} {\bibinfo {author} {\bibfnamefont {C.}~\bibnamefont
  {Froggatt}}\ and\ \bibinfo {author} {\bibfnamefont {H.}~\bibnamefont
  {Nielsen}},\ }\href
  {https://doi.org/https://doi.org/10.1016/0550-3213(79)90316-X} {\bibfield
  {journal} {\bibinfo  {journal} {Nuclear Physics B}\ }\textbf {\bibinfo
  {volume} {147}},\ \bibinfo {pages} {277 } (\bibinfo {year}
  {1979})}\BibitemShut {NoStop}%
\end{thebibliography}%

\end{document}